\documentclass[10pt,twocolumn]{revtex4}
\usepackage{graphicx}
\usepackage{psfig}
\newcommand{\be}{\begin{equation}}
\newcommand{\bea}{\begin{eqnarray}}
\newcommand{\ee}{\end{equation}}
\newcommand{\eea}{\end{eqnarray}}
\newcommand{\eps}{\epsilon}

\begin{document}

\title{Dynamics of  quantum phase transitions in Dicke and Lipkin-Meshkov-Glick models}
\author{A.P. Itin$^{1,2}$ and  P. T\"{o}rm\"{a}$^1$ }
\address{
$^{1}$Department of Applied Physics, Aalto University, P.O. Box
15100, 00076, Finland \\ $^{2}$Space Research Institute, RAS,
Profsoyuznaya str. 84/32, 117997 Moscow, Russia }

\begin{abstract}
We consider dynamics of Dicke models, with and without
counterrotating terms, under  slow variations of parameters which
drive the system through a quantum phase transition. The model
without counterrotating terms and sweeped detuning is seen in the
contexts of a many-body generalization of the Landau-Zener model
and the dynamical passage through a second-order quantum phase
transition (QPT). Adiabaticity is destroyed when the parameter
crosses a critical value. Applying semiclassical analysis based on
concepts of classical adiabatic invariants and mapping to the
second Painleve equation (PII), we derive a formula which
accurately describes particle distributions in the Hilbert space
at wide range of parameters and initial conditions of the system.
We find striking universal features in the particle distributions
which can be probed in an experiment on Feshbach resonance passage
or a cavity QED experiment. The dynamics is found to be crucially
dependent on the direction of the sweep.  The model with
counterrotating terms has been realized recently in an experiment
with ultracold atomic gases in a cavity. Its semiclassical
dynamics is described by a Hamiltonian system with two degrees of
freedom. Passage through a QPT corresponds to passage through a
bifurcation, and can also be described by PII (after averaging
over fast variables), leading to similar universal distributions.
Under certain conditions, the Dicke model is reduced to the
Lipkin-Meshkov-Glick model.
\end{abstract}
\maketitle

\section{Introduction}

Adiabatic invariance is a central issue in both quantum and
classical mechanics.  Magnitudes that we call now "adiabatic
invariants" appeared for the first time in works of L.Bolzmann
(see, e.g., \cite{AKN}). The term "adiabatic invariant" itself was
introduced by P. Ehrenfest \cite{Navarro,Ehrenfest}. Study of
changes in the adiabatic invariants  and their relation to quantum
nonadiabatic transitions was started, presumably, by P.A.M. Dirac
\cite{Dirac}; early attempts to formulate and prove quantum
adiabatic theorem were done on the eve of quantum mechanics
\cite{Born,Fermi}. Since then, destruction of adiabaticity in
quantum and classical contexts has been subject of research for
decades, with recent explosive revival of interest to that theme
due to experiments with ultracold quantum gases \cite{exp1,Jin}.


For a single-particle quantum system, very often nonadiabatic
dynamics can be described within the exactly solvable Landau-Zener
model (LZM) \cite{LZ}, where a probability of transition from an
initially occupied instantaneous ground state level to the excited
one is exponentially small in the parameter describing the
sweeping rate of the detuning between the levels.

Ultracold quantum gases experiments  may involve macroscopically
large numbers of particles, making semiclassical ({\em SC})
treatments justified (see, e.g.,
\cite{Cirac,Philips,PhysD,Polkovnikov}).
\begin{figure*}
\includegraphics[width=170mm]{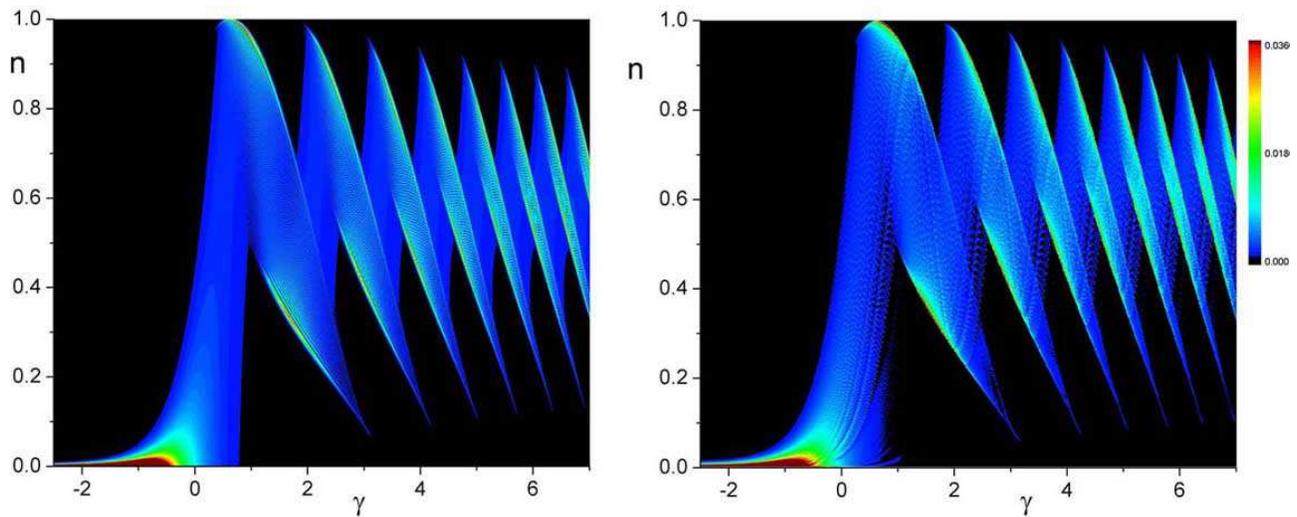}
 \caption{The time evolution of the distribution of the rescaled boson number $n$ at a "forward" sweep through the resonance.
 (a) Numerical quantum calculation with N=500 (according to Eqs.\ref{Dicke},\ref{Matrix}).
 (b) Corresponding ensemble of classical trajectories; the sweeping rate is $\eps=0.3.$} \label{Fig1}
\end{figure*} Nonadiabatic dynamics in {\em SC} models of many-particle systems
has been intensively discussed recently (see
\cite{Polkovnikov,ITorma} and references therein). In many
treatments, it was found that exponential LZM-type behaviour for
the transition probabilities is replaced with power-laws of the
sweeping rates.

A universal and accurate method for describing nonadiabatic {\em
SC} dynamics of such many-particle quantum systems was presented
by us recently \cite{ITorma}. The method is based on (i) replacing
a quantum state with ensemble of classical trajectories
distributed according to a Wigner function \cite{Wigner} of the
initial state (ii) propagating this ensemble and (iii) mapping its
dynamics near a bifurcation to a Painleve equation. For (i),(ii),
we follow Altland et al \cite{Polkovnikov} (in the present paper
we also introduce certain modifications coming from SU(2) Wigner
function formalism for spin systems \cite{ASchleich,Klimov}) .
Stage (iii) is not trivial. If initial state is chosen to be far
from the ground state, then initial classical actions are not very
small, and general separatrix crossing theory can be applied to
calculate the deviation from adiabaticity. For small initial
actions, mapping to a Painleve equation is necessary.

Thus, an interesting relation between many-particle Landau-Zener
models, Wigner functions , dynamics of quantum phase transitions
and {\em Painleve equations} \cite{Painleve} was established. The
Painleve equations have found many applications in quantum field
theory and other areas of physics (for example, they describe
correlation functions of certain fermion models \cite{Rieman}).
They possess a property that makes them in some respect similar to
linear differential equations: the Painleve property (also called
Fuchs-Kovalevskaya-Painleve property in Russian literature). That
is, their solutions have only simple poles as movable (dependent
on initial conditions) singularities in the complex time plane.
Since works of Kovalevskaya \cite{Kovalevskaya}  this property is
known to be useful to find integrable systems. Using the method of
isomonodromic deformations, it is possible to construct
asymptotics of the Painleve equations and {\em connection
formulae} (which we utilize in the present paper, following
\cite{Mitya}) by a modification of the WKB method for linear
differential equations \cite{Rieman}.

In this article, we elaborate the method \cite{ITorma} further,
implementing it for the Dicke model \cite{Dicke} with
counterrotating terms and dissipation, and Lipkin-Meshkov-Glick
(LMG) model \cite{Lipkin}, and show that peculiar properties of
PII can appear in experiments with cold atoms.  LMG model arises,
in particular, as an effective model at certain range of
parameters in an experiment recently done by the group of
T.Esslinger \cite{Esslinger,Keeling}. Other experimental
realizations of LMG model were also suggested \cite{Larson}. On
the basis of Dicke model with counterrotating terms we confirm
that the method works for systems whose classical counterpart is a
few-dimensional Hamiltonian system experiencing a pitchfork
bifurcation at a critical value of a parameter.

In the next section, we describe the Dicke model without
counter-rotating terms and its connection to applications.

Section III considers a passage through the quantum phase
transition in the LMG model. We note, following Keeling et al.
\cite{Keeling}, that in the Dicke model {\em with counterrotating
terms} under conditions relevant to the experiment
\cite{Esslinger}, LMG model arises as  an effective model after
adiabatic elimination of the photon field. Dynamics of quantum
phase transitions is studied, and the universal distributions
after the sweep of the parameter are found.

Section IV considers Dicke model with counterrotating terms
\cite{Dicke} and comparable frequencies of the photon field mode
and the two-state system (i.e., separation of timescales of
Section III allowing mapping to LMG model does not happen here).
In this model, a quantum phase transition between normal and
superradiant phases was studied \cite{Brandes}. Also, signatures
of quantum chaos were investigated \cite{Brandes}. We analyze the
{\em dynamics} of the quantum phase transition. The corresponding
classical Hamiltonian system has two degrees of freedom; this, in
particular, leads to chaotization of large area of phase space in
the "superradiant" phase. Naively, it should preclude the
application of our method. However we demonstrate that the passage
through a bifurcation happens in a one-dimensional way and can be
also described by PII. For not very fast sweeps, most of the
classical wavepacket will remain in the vicinity of new
equilibria. There, the phase space is regular: one may introduce
classical actions and modify the method of Sections II-III
correspondingly.

Section V presents the conclusions.

\section{Non-adiabaticity of a many-particle Landau-Zener problem: forward
and backward sweeps  }

\subsection{The time-dependent Dicke model without counterrotating terms as a many-particle LZM}

We consider the following many-particle LZM: time-dependent Dicke
model \cite{Dicke} which has numerous applications in quantum and
matter-wave optics \cite{Altman, PhysD, Polkovnikov}. The model
can be written as \be \hat{H} = -\frac{\gamma(t)}{2} \hat
b^\dagger \hat b + \frac{\gamma(t)}{2} \hat S^z +
\frac{g}{\sqrt{N}}(\hat b^{\dagger} \hat S^- + \hat b \hat S^+),
\label{Dicke} \ee where $\frac{g}{\sqrt{N}}$ is the coupling
strength, $\hat S^{\pm} = \hat S_x \pm i \hat S_y $ are spin
operators, $\hat b^\dagger$ and $\hat b$ are creation and
destruction operators of a bosonic mode, $\gamma(t)= \pm 2\eps t$
is the detuning from the resonance, and $\eps$ is the sweeping
rate of the bosonic mode energy. The spin value $S$ can be
considered macroscopically large $S=N/2 \gg 1$; usually the
physical origin of the effective spin variable $S$ is a collection
of two-level systems (spin-1/2 particles). Among many possible
applications, let us mention the Feshbach resonance passage
\cite{Jin,Timmermans,JavanainenD,Altman, Gurarie2, Kohler},
dynamics of molecular nanomagnets \cite{Garanin}, and cavity QED
with Bose-Einstein condensates (BEC) \cite{Keeling,Polkovnikov}.

With $N=1$, one recovers the standard Landau-Zener model. In the
context of a Feshbach resonance passage in a Fermi gas, an
equivalent realization of the Hamiltonian (\ref{Dicke}) is $$ \hat
H =  \frac{\gamma(t)}{2} \sum_{i=1}^N \left( \hat n_{i \uparrow}+
\hat n_{i \downarrow} \right) + \frac{g}{\sqrt{N}}\sum_{i=1}^N
\left( \hat b^\dagger \hat c_{i\downarrow} \hat c_{i\uparrow} +
h.c. \right),
$$
where $\hat c_{i,\sigma}$ and $\hat b$ are the fermion and boson
annihilation operators, respectively, $\hat n_{i\sigma} = \hat
c_{i\sigma}^\dagger \hat c_{i \sigma} $, and $\sigma=\uparrow,
\downarrow$. Coupled atom-molecular BECs are described by a
similar model whose dynamics in the limit of large $N$ becomes
equivalent to (\ref{Dicke}) with the replacement $\gamma \to
-\gamma$ \cite{Polkovnikov,ITorma}. That is, in the thermodynamic
limit of the degenerate model (\ref{Dicke}) association of
fermionic atoms becomes near equivalent to dissociation of a
molecular BEC, and vice versa. We  discuss two driving scenarios
here: "forward" and "backward" sweeps.

In the "forward" sweep, starting in the distant past with some
small initial number of bosons ${N_b(t)|}_{-\infty} \equiv \langle
\hat b^\dagger \hat b \rangle (t)|_{-\infty} =N_- \equiv
{\bf{n_-}} N$, we want to calculate the final number of bosons
$N_b(t)|_{+\infty} \equiv {\bar n} N$ {\em and} its distribution
$P( \bar n)$ as a function of the sweeping rate $\eps$ and the
initial bosonic fraction ${\bf n_-}$. To be more specific, in the
infinite past we start in a (Dicke) eigenstate of (\ref{Dicke}):
$|\psi_{-\infty} \rangle= |N_- \rangle |S,S_z \rangle = |N_-
\rangle |\frac{N}{2},\frac{N}{2}-N_- \rangle$, where $N_-$ can be
zero in case we start at the ground state.   We consider the
sector $S=N/2$ for clarity, but other values of $S$ can be treated
analogously (Hilbert space of (\ref{Dicke}) is decoupled on
sectors with definite total spin, or "cooperation number", $S$).
Matrix elements of (\ref{Dicke}) have the form
 \bea H_{n,n'} &=& - \gamma(t) \delta_{n,n'} + n
\delta_{n,n'+1}\sqrt{N-n'} /\sqrt{N}  \label{Matrix}\\ &+& n'
\delta_{n,n'-1}\sqrt{N-n} /\sqrt{N} \nonumber \eea (we use $g=1$
for convenience in this paper, which can always be achieved by
rescaling of time), therefore the model can be referred to the
class of generalized Landau-Zener models \cite{Shytov} in the case
of linear driving $\gamma(t) \sim t$.

In the second dynamical scenario ("backward"), we start in the
ground state of (\ref{Dicke}) at large positive value of $\gamma$
and make an "inverse" sweep to large negative $\gamma$ (see also
\cite{ITorma}). This scenario is relevant to association of Bose
atoms to a molecular BEC \cite{Polkovnikov}. On a purely classical
level, there is a drastic asymmetry in the forward and backward
sweeps. Starting in the classical ground state of (\ref{NLZM}) at
$ \gamma=-\infty$ (i.e. $n=0$, "all-atom" mode in context of Fermi
association), one obtains a situation where dynamics is absent: a
phase point always remains in the equilibrium. To trigger
dynamics, one needs either to consider nonzero initial population
of the molecular mode (as done, in particular, in \cite{ PhysD}),
or carefully take into account quantum fluctuations
\cite{Polkovnikov}. This should be contrasted with the "backward"
sweep \cite{ITorma}, where, starting in the classical ground state
at $\gamma = +\infty $ (i.e., $n=1$, "all-atom" mode in context of
Bose atoms association and "all-molecule" mode in context of Fermi
association), there will be dynamics, and the system can be
considered on a purely classical level \cite{Polkovnikov,ITorma}.
Such asymmetry in classical dynamics with respect to direction of
sweeps is preserved on a semiclassical level as well, as will be
shown below.

The Dicke model \cite{Dicke} and its equivalent realizations have
been studied already using various techniques. Diagrammatic
methods employed in \cite{Polkovnikov} work well for small $N$ but
do not allow to get close to the {\em SC} limit for large $N$.
Another approach is a {\em SC} treatment based on classical
adiabatic invariants (see \cite{PhysD,Polkovnikov} and references
therein). At $|t|=\infty$ and $N \to \infty$ the relative number
of bosons $N_b/N$ corresponds to a classical action $I$ of an
effective Hamiltonian system, so that for large but finite $N$ the
problem can be mapped to the calculation of a change of classical
actions of a properly prepared ensemble of trajectories. It is
inspiring that calculations with ensembles of classical
trajectories (defined below) reproduce full quantum calculations
very well (see Fig.\ref{Fig1}).

\subsection{Classical ensembles and quantum dynamics}
 \subsubsection{Classical limit of the system}
The classical limit of (\ref{Dicke}) can be obtained from the
Heisenberg equations of motion by factoring all operator products:

\bea
\dot{x} &=& -\frac{\gamma}{2}  y - g z p, \nonumber\\
\dot{y} &=&\frac{\gamma}{2}  x - g z e, \nonumber\\
\dot{z} &=& g (x p + y e), \label{DickeCl} \\
\dot{e} &=& -\frac{\gamma}{2} p -g y,  \nonumber\\
 \dot{p} &=&  \frac{\gamma}{2} e - g x   \nonumber,\\
 \eea
where the variables $x,y,z$ are components of the Bloch vector,
while $p,e$ are the components of the boson ("radiation") field:

\bea p = \frac{i}{\sqrt{N}}(b^* &-& b), \quad
e=\frac{1}{\sqrt{N}}(b+b^*), \nonumber\\
 x,y,z &=& \frac{2}{N} S_{x,y,z}. \eea
The length of the Bloch vector $x^2+y^2+z^2=1$ is the integral of
motion. The equations (\ref{DickeCl}) are equivalent to the
Hamiltonian equations of motion of the following Hamiltonian:
 \be
H= \frac{\gamma}{2}(z-I)-g\sqrt{2I} \sqrt{1-z^2}\sin(\theta-
\phi),  \label{HamRWA} \ee where $(z,\theta)$ and $(I,\phi)$ are
canonically conjugated pairs of variables related to the variables
of (\ref{DickeCl}) by
\bea x &=& \sqrt{1-z^2} \cos \theta, \nonumber\\
y &=& \sqrt{1-z^2} \sin \theta, \\
p &=& \sqrt{2 I} \cos \phi, \nonumber\\
e &=& \sqrt{2 I} \sin \phi. \nonumber \eea

Eqs.(\ref{DickeCl}) posses an additional integral of motion:
$\tilde L =I+z$, which makes the Hamiltonian (\ref{HamRWA}) at
constant $\gamma$ integrable.  Switching from the variables
$(I,\phi)$, $(z,\theta)$ to $(\tilde L=I+z,\tilde \theta=\theta)$,
$(\tilde n = I, \tilde \phi = \phi-\theta)$, one obtains $H= -
\gamma \tilde n + g \sqrt{2 \tilde n } \sqrt{1-(\tilde L-\tilde
n)^2} \sin \tilde \phi.$ Denoting $\tilde n= 2n $, $1-\tilde L =
2L$, shifting the phase $\tilde \phi = \phi - \pi/2$, rescaling
$H$ by a factor of 2 and setting $g=1$ one obtains
 \be H= -\gamma n -2 \sqrt{ n(n+L)(1-L-n)} \cos \phi,
\label{PolkovnikovF} \ee where $n \in[0,1]$ corresponds to the
rescaled number of bosons $N_b/N$, and $\phi$ is the canonically
conjugated phase.

\subsubsection{Classical distribution}


Distributions of $n, \phi$ as well as distribution of integral of
motion $L$ follow from the initial Wigner function of the system.
There are at least two different ways to define Wigner function of
the system: using decomposition of spin on Schwinger bosons leads
to highly oscillatory Wigner function from which an effective
positive distribution should be derived \cite{Polkovnikov,ITorma},
while implementing SU(2) Wigner function for spin system
\cite{ASchleich,Klimov} immediately gives positive Wigner
distribution of the initial state because both eigenstates of
$\hat S_z$ with maximal absolute values of $S_z$ , $|\pm S,S
\rangle$ posses positive, non-oscillatory Wigner function.
Interesting enough, we found that in the limit of large $N$ both
approaches give the same results: SU(2) Wigner function  leads to
the same exponential distributions as that used in
\cite{Polkovnikov,ITorma}. Indeed, as shown in \cite{Klimov} the
eigenstate $ |\frac{N}{2},\frac{N}{2} \rangle$ has SU(2) Wigner
function $W_{\frac{N}{2}}(\Theta) \approx \cos^N \Theta [1+\cos
\Theta]= z^N (1+z)$ (see Appendix), where $\Theta$ is an angle on
Bloch sphere: $z=\cos \Theta$. This function at large $N$ is
asymptotically close to exponential: $z^N(1+z) \approx 2
\exp[-(1-z)N]$.

On the other hand, the approach used in \cite{ITorma} is as
follows. The classical limit of (\ref{Dicke}) can be also obtained
by introducing Schwinger bosons $\hat \alpha, \hat \beta: \quad
\hat S^+ = \hat\alpha^\dagger \hat\beta, \quad \hat S^- = \hat
\beta^\dagger \hat \alpha, \quad \hat S_z= (\hat n_{\alpha}-\hat
n_{\beta})/2,$ and using the $c-number$ formalism
\cite{Polkovnikov,PhysD}. The resulting classical system has 3
pairs of variables ($n_{\alpha,\beta,b}$ and conjugated phases
$\theta_{\alpha,\beta,b}$) and 2 integrals of motion (${\cal
N_{\alpha}} \equiv n_{\alpha}+n_{\beta}, \quad L \equiv n_{\beta}-
n_{b} $). To obtain the Wigner function, one notes that far from
the resonance, the quantum system is effectively decoupled on 3
oscillators, 2 of them being in the ground state, and the third
one in the high-lying Fock state $|N \rangle$. With a good
accuracy, $W(n_{b,\alpha,\beta}) \sim \exp(-2 n_b-2
n_{\beta})\delta(n_{\alpha}-1)$. The origin of such distribution
is as follows. Wigner function of the ground state of a oscillator
is $W_0= 2 \exp(-2n_{b,\beta})$. Wigner function of a highly lying
Fock state is highly oscillatory and, in principle, is not
suitable for preparation of a classical distribution we need.
However, the effective smooth distribution (Gaussian) was derived
by Gardiner et al. \cite{TWAG}, and it is much narrower than the
distribution of the ground state. Therefore, with a good accuracy
it can be replaced with the delta-function, as authors of
\cite{Polkovnikov,TWAP} did. In all distributions mentioned above,
phases $\theta_{\alpha,\beta,b}$ are uniformly distributed on
$[0,2\pi]$. It means initial phase $\phi$ is also uniformly
distributed on $(0,2\pi)$; also, besides exponential distribution
of initial $n$, we have to take into account distribution of the
parameter $L$.


\subsubsection{Nonadiabatic dynamics of classical trajectories}

Let us now analyze the classical trajectories from the initial
distribution. Consider firstly the case $L=0$, then the
Hamiltonian reduces to \be H = -\gamma n - 2 n\sqrt{1-n}\cos \phi.
\label{NLZM}\ee Note that by means of a trivial change of
variables \cite{footnote}, the Hamiltonian becomes the same as
analyzed e.g. in \cite{PhysD}.  The classical phase space of the
Hamiltonian (\ref{NLZM}) at fixed values of the parameter $\gamma$
is described in \cite{PhysD,ITorma}. If $\gamma <-2$, there is
only one stable elliptic point on the phase portrait (Fig.2a). At
$\gamma = -2$, a bifurcation takes place. There are two saddle
points at $n = 0, \cos\phi=-\gamma/2$, and a newborn elliptic
point at $\phi = 0$. The trajectory connecting these two saddles
(the separatrix) separates rotations and oscillating motions. At
large positive $\gamma$, again there is only one elliptic
stationary point at $\phi = 0$, and $n$ close to $1$. The
classical action is defined as in \cite{PhysD,Polkovnikov} and is
shown graphically in Fig. 3: shaded areas, divided by 2$\pi$. At
$\gamma=-\infty$, the action coincides with $n$: $I=n$, and
canonical "angle" variable $\varphi$ coincides with $\phi$. At
$\gamma=+\infty$, we have $I=1-n$. We are interested in small
initial actions, corresponding to the ground state of the quantum
system.

\begin{figure}
\includegraphics[width=80mm]{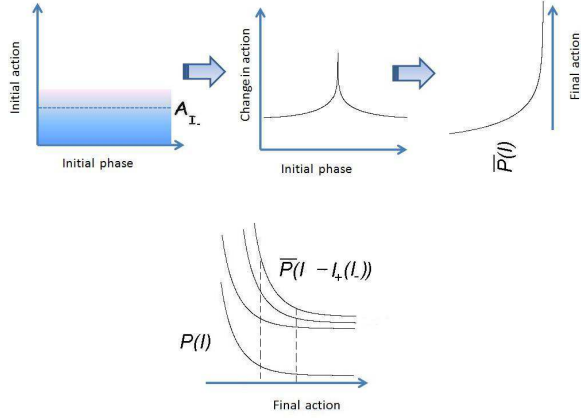}
\caption{ Build-up of distributions in the action space from the
initial Wigner distribution and the change in action during the
sweep through the phase transition. This schematic picture applies
(with certain modifications) to all three models considered in
Sections II-IV. The top left figure illustrates initial
distribution $W(I_-) \sim \exp(-N I_- )$, which is exponential in
the initial action and uniform in the initial phase. From that
distribution, we firstly single out a narrow slice ${\cal
A}_{I_-}$ of phase points with uniform distribution of initial
phases and nearly constant action $I_-$. As explained in the text,
phase points from that slice experience a change in the action
$\Delta I(I_- )$ which can be decomposed on the phase independent
$(I_+ (I_-) \sim \eps \ln \pi I_-/\eps )$ and the phase-dependent
(${\cal I}_+(\xi) \sim \ln 2\sin \pi \xi $ ) parts (where $\xi
\in(0,1)$ is the pseudophase proportional to the initial phase for
the case of small initial actions, as explained in the text).
Importantly, the phase-dependent part of the change in the action
do not depend on $I_-$. This phase-dependent part of the change in
the action is illustrated in the top center figure. It maps the
uniform distribution in the initial phase to a distribution $\bar
P(I)$ in action space, as illustrated in top right figure.  That
is, $\bar P(I) \sim \frac{1}{\sqrt{4\exp\left(\pi I \right)-1} }.$
Now, to find the overall distribution in action space one needs to
integrate over all values of $I_-$, i.e. to sum up contributions
from all ${\cal A}_{I_-}.$  As illustrated in the bottom figure,
the final distribution is obtained as $ P(I) = \int dI_-
W(I_-)\bar{P}(I-I_+(I_-))$, see also the corresponding discussion
in Section III. } \label{illustr}
\end{figure}
\begin{figure}
\includegraphics[width=80mm]{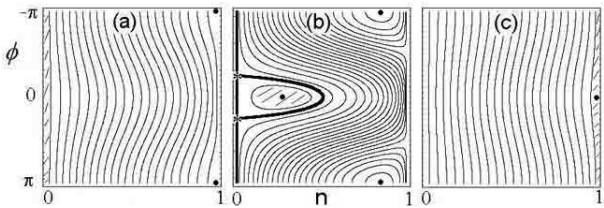}
 \caption{Phase portraits of (\ref{NLZM}). From (a) to (c):
 $\gamma=-10,-1.4,20$ correspondingly. Saddle points are denoted by asterisks,
the bold line is the separatrix. Shaded areas illustrate definitions of the
classical actions (see text).} \label{Fig2}
\end{figure}

We split the initial classical ensemble in slices with equal
actions $I_-$ (let us call such an ensemble
 ${\cal A_{I_-}}$), and analyze dynamics of each slice in detail.
Fig.(\ref{illustr}) illustrates the algorithm of the analysis.
 In the infinite past, in variables $n,\phi$ the slice ${\cal
A_{I_-}}$ corresponds to a stripe $n=const, \phi \in [0,2\pi]$
(see Figs.(\ref{illustr},\ref{Fig2}). Initial actions $I_-$ of a
classical ensemble in the case of a fully polarized $({\bf n_-}
=0)$ initial quantum state are of the order $\frac{1}{N}$ due to
quantum fluctuations. In case there is some non-zero initial
population of the bosonic mode, $I_-$ can be much larger than
$\frac{1}{N}$. There are three small parameters in the model:
$\frac{1}{N}$, the sweeping rate $\eps,$ and the initial action
$I_-$. Naturally, $\frac{1}{N}\ll \eps$. Still, freedom of
choosing a value of the ratio $I_-/\eps$ remains. Assuming $I_-
\gg \eps$ and $\eps,1/N,I_- \ll 1$, which corresponds to
considerable initial population of the bosonic mode (see Fig.4),
Ref.\cite{PhysD} applied an approach based on the separatrix
crossing theory \cite{Cary,N86,AKN}. From the experimental point
of view, however, it is important to analyze also cases of very
small initial actions.

\begin{figure}
\includegraphics[width=80mm]{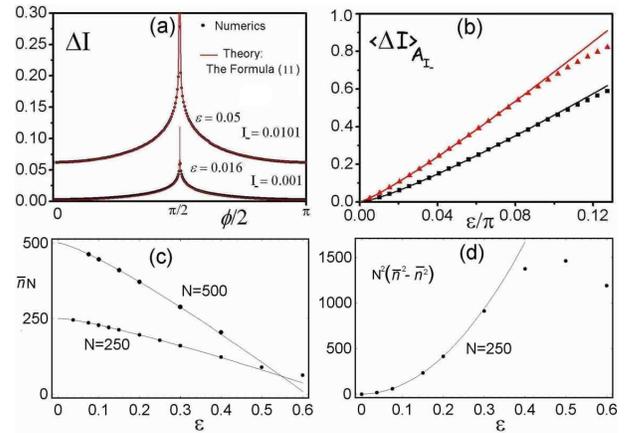}
 \caption{
(a) Change in the action $\Delta I$ of classical trajectories from
the ensemble ${\cal{A}_{I_-}}$ as a function of the initial phase
(Eq.(\ref{Formula})) . The change $\Delta I$ describes the
deviation from adiabaticity. We shift all curves horizontally so
that their maxima are at $\pi/2$. We found the change in the
action is highly phase-dependent. (b) Average change of the
adiabatic invariant $\langle\Delta I \rangle_{{\cal{A}}_{I_-}}$ as
a function of the sweep rate $\eps$.
 Lines are the theoretical curves (Eq.(\ref{meanDeltaI})). Squares, triangles are numerics for $I_-=10^{-3},10^{-4}$,
 respectively. (c) Number of created bosons from quantum calculations with $N=250$ and $N=500$ (dots); from the Formula
 (\ref{final}) obtained by averaging over the initial
 distribution $W \sim \exp[-2n_b-2n_{\beta}]$ (solid lines). (d) Dispersion of the number
of bosons for N=250; dots: quantum calculations, solid line: the
formula (\ref{final}), i.e. $N^2(\bar{n}^2-\bar{n^2})=\frac{N^2
\eps^2}{6}$. } \label{Fig3}
\end{figure}

\begin{figure}
\includegraphics[width=78mm]{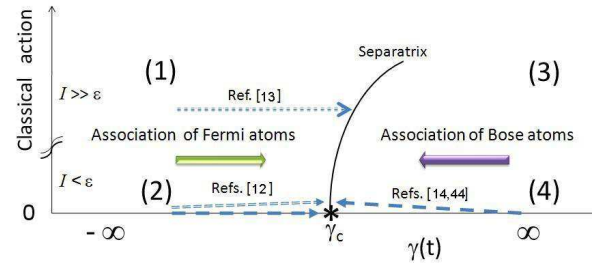}
 \caption{Regimes of initial conditions and directions of sweeps for a
nonlinear LZM Eq.(\ref{NLZM}). Regimes "1" and "3" correspond to
starting far above the classical ground state, where the
separatrix crossing theory can be applied \cite{PhysD}, which
results in Eq.(\ref{PhysDD}). Regimes "2" and "4" can be described
by mapping to PII derived here. Starting in the classical ground
state in regime 4 (i.e., in "all-atoms" mode in context of bosonic
association, or "all-molecules" mode in fermionic association) was
considered, e.g., in \cite{ITorma,Ishkhanyan}. A linear power-law
with the coefficient $\frac{\ln 2}{\pi} $ was derived in
\cite{ITorma}. Starting in the ground state in regime 4, the
classical system Eq.(\ref{NLZM}) always remains in the
equilibrium. Taking into account quantum fluctuations of the
ground state of Eq.(\ref{Dicke}) \cite{Polkovnikov} produces an
ensemble of initial conditions which leads to Eq.(\ref{Formula}).
} \label{Fig4}
\end{figure}

Our method for the regime of small initial actions is as follows.
To calculate the deviation from adiabaticity, we note that most of
the change of the classical action of a phase point happens as it
travels near the separatrix and, especially, near the saddle
points that arise during the bifurcation as $\gamma$ reaches
$-2_-$ \cite{PhysD} (in our new variables, the arising saddle
point is located in the origin). Close to the saddle point $n$ is
small and one can expand $\sqrt{1-n}$ in series. Let us introduce
the variables
 $$P=2\sqrt{n} \cos{\frac{\phi}{2}},\quad Y=2\sqrt{n} \sin{\frac{\phi}{2}}.$$ Near
the bifurcation, at $\gamma \approx -2$, we therefore get an
effective Hamiltonian $$H=-\frac{P^2}{2} (\gamma/2 + 1) -
\frac{Y^2}{2} (\gamma/2-1) + \frac{P^4}{16},$$ where higher order
terms in $Y$ and $P$ have been neglected. We neglect also the
time-dependence of the coefficient $\gamma/2-1$ of the second
term. Then, shifting the origin of time, we obtain the Hamiltonian
$-\frac{P^2}{2} \eps t + Y^2+ P^4/16 $. After simple rescalings
$P= 2^{4/3} P'$, $Y=2^{2/3}Y' $,$t=t'/2^{1/3}$, $H= 2^{7/3} H' $,
the Hamiltonian becomes $$H=-\frac{P^2}{2}\eps t +
\frac{Y^2}{2}+\frac{P^4}{2}$$ (primes over new variables omitted).
We replace also momenta and coordinates: $P \to Y, \quad Y \to
-P$. Let us now introduce a rescaling transformation that makes
the essential mathematics of the problem as clear as possible: $$
Y=\eps^{1/3} \tilde Y, \quad P= \eps^{2/3} \tilde P, \quad t=
\eps^{-1/3} s, \quad H = \eps^{4/3}\tilde H. $$ The Hamiltonian
becomes (omitting tildes over new variables) $$ H= -s
\frac{Y^2}{2} + \frac{P^2}{2} + \frac{Y^4}{2}.$$ This Hamiltonian
does not have a small parameter any more, and the loss of
adiabaticity is evident. An important property of the bifurcation
we are considering is that the effective Hamiltonian leads to the
second Painlev\'{e} equation (PII) \be \frac{d^2 Y}{ds^2}= sY - 2
Y^3. \label{Penleve} \ee Asymptotics of PII were investigated by
Its and Kapaev \cite{Its} (see also \cite{Mitya}) using a method
of isomonodromic deformations \cite{Novokshenov}. At $s \to
-\infty$ the asymptotic solution to (\ref{Penleve}) is
\cite{Its,Mitya} $$ Y(s)= \alpha (-s)^{-\frac{1}{4}} \sin \left(
\frac{2}{3}(-s)^{3/2} + \frac{3}{4} \alpha^2 \ln(-s) + \phi
\right),$$ and in the limit $s \to + \infty$ it is
 $$
Y(s) = \pm \sqrt{\frac{s}{2}}  \pm \rho(2s)^{-\frac{1}{4}} \cos
\left( \frac{2 \sqrt{2}}{3} s^{3/2} - \frac{3}{2} \rho^2 \ln(s) +
\theta \right),$$ where $(\alpha, \phi)$ and $(\rho,\theta)$ are
the "action-angle" variables characterizing the solutions in the
limits $s \to \pm \infty$. As $s \to \pm \infty $, the adiabatic
invariant $I_p$ of equation (\ref{Penleve}) approaches the
quantities $I_p^-$ or $I_p^+$ which are defined (to the lowest
order terms) as $$I_p^-=\frac{\alpha^2}{2}, \quad I_p^+
=\frac{\rho^2}{2}.$$ The jump in the adiabatic invariant $\Delta
I_p = 2I_p^+ - I_p^-$  can be found from general relations between
$\rho^2$ and $\alpha^2$ as $$ I_p^+ = \frac{1}{2\pi} \ln
\frac{1+|p|^2}{2 |\mbox{Im}(p)|}, \quad p = \sqrt{e^{2\pi
I_p^-}-1} e^{[\tilde f(I_p^-)-i \phi ]} $$ (the function $f(I_p^-)
$ is not important for our discussion, see e.g.
\cite{Its,Mitya,ITorma}). Returning back to the original variables
and the Hamiltonian (\ref{NLZM}), we get the Formula: \be \Delta I
=  \eps \Bigl( \frac{I_-}{\eps} - \frac{2}{\pi} \ln \sqrt{\exp
\left[ \frac{\pi I_-}{\eps} \right]-1} -\frac{2}{\pi} \ln (2 \sin
\pi \xi) \Bigr), \label{Formula} \ee where for the ensemble ${\cal
A}_{I_-}$ one has $\pi \xi =(f(I_-)+\frac{ \phi}{2})$; $\xi$ is a
quasi-random variable uniformly distributed on (0,1)
\cite{explanation,probabilistic}. The formula predicts the average
change in the action to be \be \langle \Delta I \rangle_{{\cal
A}_{I_-}} = \eps \Bigl( \frac{I_-}{\eps} - \frac{2}{\pi} \ln
\sqrt{\exp \left[ \frac{\pi I_-}{\eps} \right]-1} \quad \Bigr),
\label{meanDeltaI} \ee  which means that the final number of
bosons (in the ensemble ${\cal A}_{I_-})$ is
$$ \bar n = 1 - I_- - \langle \Delta I \rangle_{{\cal A}_{I_-}}.
$$ It predicts also the final distribution $P(\Delta I)=P(1-I_- -
\bar n)$. All moments $M^k_{{\cal A}_{I_-}} \equiv \langle (\Delta
I- \langle\Delta I \rangle)^k \rangle_{{\cal A}_{I_-}} $ are easy
to calculate, for instance $$ M^2_{{\cal A}_{I_-}} = \left(\frac{2
\eps}{\pi} \right)^2 \int_0^1{d \xi \ln^2 \large( 2 \sin \pi \xi
\large) } = \frac{\eps^2}{3}. $$

Note that the phase-dependent part in (\ref{Formula}) exactly coincides with the
result of \cite{PhysD}, obtained under the different conditions
$\eps \ll I_- \ll 1$. Moreover, if $I_- \gg \eps$, we recover this result of
\cite{PhysD}, which in the present variables is \be \Delta I =
-\frac{2\eps}{\pi} \ln( 2\sin \pi \xi ), \label{PhysDD} \ee where
the pseudophase $ \xi \in (0,1)$ is a quasi-random variable. Such
a change in the action has zero mean value, nevertheless it
introduces spreading in particle distribution since $\langle
\Delta I^2 \rangle \sim \eps^2$.  When $I_- \ll \eps$, we have a
qualitatively different result which resembles that of
\cite{Polkovnikov} (i.e.\ $ I_+ = \eps \ln I_-/\pi$, with
$I_-=\frac{1}{N}$): \be \Delta I = -\eps \Bigl(
\frac{1}{\pi} \ln \left( \frac{\pi I_-}{\eps} \right)
+\frac{2}{\pi} \ln (2 \sin\pi \xi ) \Bigr). \label{Formula2} \ee

Formula (\ref{Formula}) can be used over a wide range of values of
$\eps$ and $I_-$ (we require $I_- \ll 1$ and $\eps |\ln I_-|
\lesssim 1$ ). Qualitatively, it is important that the final
distributions are determined not only by the average change in the
action (\ref{meanDeltaI}), but also by the phase-dependent part
(\ref{PhysDD}), which is therefore important in {\em amplification
of quantum fluctuations}. The profile of the phase-dependence of
$\Delta I$ has a striking universality: we found that several
other infinitely-coordinated models, e.g. LMG model \cite{Lipkin},
have analogous behavior during linear sweep of a control parameter
through a critical value. The comparison between the numerical and
analytical results for $L=0$ are given in Figs. \ref{Fig3}a,b,
where predictions of the Eq.(\ref{Formula}) start to deviate from
the classical numerics only at large $\eps$ (such that
$-\frac{\eps}{\pi} \ln \frac{\pi I_- }{\eps} \sim 1$).

Now let us remind that the full classical ensemble, governed by
the Hamiltonian (\ref{PolkovnikovF}), has distribution of
integrals of motion $L$; i.e., analysis so far dealt only with a
subspace of the initial conditions. To take into account this
distribution, we consider slices of the full classical ensemble
with equal values of $ n +\frac{L}{2} \equiv x$ and uniform
distribution of $L \in (-2x, 2x)$; averaging over $\phi$ and $L$
within the slice, we found that the formula (\ref{meanDeltaI})
(derived for the "central" point of the slice, i.e. $L=0,x=I_-$)
acquires additional coefficient of 2 inside the logarithm: $I_-
\to 2I_-$. The amplitude of the phase-dependent part is also
modified, although it retains its characteristic form shown in
Fig. 4a; the final result of averaging over the 3-dimensional
distribution (in $\phi,I_-,L$) of the initial ensemble of phase
points is: \bea \bar{n} &=& 1 -\frac{\eps}{\pi} \left( \ln
\frac{\eps N }{\pi} + C_{\gamma} \right), \label{finala} \\ \quad
\bar{n}^2 &-& \bar{n^2} = \frac{\eps^2}{6}, \label{final}\eea
where $C_{\gamma}$ is the Euler constant. Comparison with
numerical quantum calculations are given in Figs. 4c,d.

Let us now briefly discuss  the "backward" sweep (see also
\cite{ITorma} and Section III), whose properties can also be
derived from PII equation. The main feature of the inverse sweep
is phase-independence of the change in the action in the limit of
small $I_-$. Deviation from adiabaticity at slow sweeps strictly
follows a linear power-law (i.e., phase-dependent terms become
negligible). The coefficient of the linear power-law was estimated
in \cite{Ishkhanyan} to be equal to $\frac{2}{3\pi} \approx 0.21$.
The asymptotically exact value of the coefficient was found by us
from mapping to PII \cite{ITorma}: $\frac{\ln2}{\pi} \approx
0.2206$. The most appealing physical system to implement this
process is photo- or magneto- association of Bose condensates,
which was shown to be well described by a two-mode mean-field
model. "Forward" sweep would correspond to dissociation of BEC.
Experimentally, studying distributions after a sweep could be
achieved by making many runs at each particular sweeping rate.

\subsection{Conclusions} In summary, we have found novel universal
features in the dynamics of a many-particle LZM.
 In particular, the universal profile of Fig. 3a, partly responsible
 for broad distributions after the sweep, was found. Such universal profiles are very interesting
 from physical point of view, recall e.g. the Fano profile \cite{Fano} in AMO physics. There
 is a classical analogy of the Fano profile (which partly explains its universality):
 response of two coupled linear oscillators to periodic driving of one of them.
Analogously, when an underlying quantum system experiences a {\em
quantum phase transition} (QPT), its classical counterpart
experiences a bifurcation. Response of a semiclassical system to
slow driving which pushes it through the symmetry-breaking
bifurcation, as was shown here, leads to a universal profile $\sim
\ln |\sin \pi \xi|$ being imprinted in distributions.

We depicted schematically four different regimes  of initial
conditions and directions of the sweeps in Fig. 5. It is
interesting to compare the expression for the mean number of
produced bosons (\ref{finala}) with the results of Refs.
\cite{Polkovnikov, Gurarie2}. The main difference of our result
for the "forward" sweep from Ref. \cite{Polkovnikov} is the
coefficient $\eps$ inside the logarithm of Eq.(\ref{finala}). The
model neglects kinetic dispersion of the fermions, but takes into
account quantum effects. If we start in the ground state,
effectively we have two small parameters: $1/N$ and the sweeping
rate $\eps$. In the limit of $1/N \to 0$ the model possess a QPT.
At the same time, the model of \cite{Gurarie2} neglects quantum
effects but takes into account fermionic dispersion. It also has
two small parameters, the width of a Feshbach resonance $\gamma$
and dimensionless sweeping rate $\Gamma$. In the limit $\gamma \to
0$ the model of \cite{Gurarie2} possess a QPT. Our formula
Eq.(\ref{finala}) is very close to result of Ref. \cite{Gurarie2},
even though the dynamics is different.

The results reported here are highly relevant for accurately
describing the Feshbach resonance passage in ultracold Fermi and
Bose gases and may motivate further experimental activity on that
theme. Furthermore, we believe our method will have important
applications in the field of {\em dynamics} of QPT
\cite{QPTP,QPTZoller,DziarmagaAP,QPT}. An important and
interesting direction of the future research would be to extend
our method to higher dimensions, i.e. to consider a nonuniform
spatially extended system, with coupling of the conversion
dynamics to phonon modes, etc.

\section{Dicke Model with cavity decay and high frequency of the field mode: effective Lipkin-Meshkov-Glick  model}

\subsection{The model}

Dicke model with counterrotating terms \be H = \omega_0 J_z + \bar
\omega b^\dagger b + \frac{ \bar\lambda}{\sqrt{2J}} (b^\dagger +
b)(J_+ + J_-), \label{Dicke22} \ee considered in detail in the
next Section, can be reduced to the Lipkin-Meshkov-Glick model in
the limit of high frequency of the field mode $\bar \omega$, as
shown e.g. in \cite{Keeling2}. Consider the Hamiltonian
(\ref{Dicke22}) with added dissipation (cavity decay $\kappa$),
such that $ \bar \omega \gg \kappa \gg \omega_0$ \cite{Keeling2}.
Classical equations of motion  become

\bea
\dot{x} &=& - \omega_0 y \nonumber\\
\dot{y} &=& \omega_0 x - 2 \lambda e z  \nonumber\\
\dot{z} &=& 2 \lambda e y \label{DickeCl3}\\
\dot{e} &=& \bar \omega p  - \kappa e \nonumber\\
\dot{p} &=& -\bar \omega e - 2 \lambda x - \kappa p. \nonumber\\
\eea The boson field in the corresponding Heisenberg equations of
motion can be adiabatically eliminated:

\be b = -\frac{2i \lambda J_x}{(\kappa + i \omega) \sqrt{N}}.
 \ee
which results in the Lipkin-Meshkov-Glick (LMG) model for the spin
system \cite{Keeling2} \be H_{LMG} = \omega_0(t) J_z - \frac{\bar
\omega}{\kappa^2 + \bar\omega^2} \frac{(2 \lambda)^2}{N} J_x^2
\equiv \omega_0(t) J_z - \frac{\sigma(t)}{N} J_x^2. \ee

The LMG model was introduced as a toy model to test the quality of
approximations used in multiparticle systems. The structure of the
eigenstates of $H_{LMG}$ is compatible with Hartree Fock
solutions, therefore one can test, e.g., the validity of Random
Phase Approximation against the exact solution. It was also
studied recently in the context of dynamics of quantum phase
transitions \cite{Fazio,FazioRef,Vidal}.

We can study a sweep through a second-order quantum phase
transition by changing the coupling $\sigma(t)$, say from 0 to 2,
using a smooth pulse that is approximately linear in the vicinity
of the critical value of $\sigma$, i.e. similar to the previous
section. We can choose to change detuning $\omega_0$ instead, thus
realizing a clearer analog of the Landau-Zener model (also, the
problem becomes analogous to that studied in \cite{Fazio}). The
system is very much related to that studied in Section II, but now
the dynamics of almost all classical trajectories are captured by
PII and we can study quantum-classical correspondence in a greater
detail.

After the rescalings $H \to H/\sigma, t \to t/\sigma$ Hamiltonian
can be brought to the form \be H= -\gamma(t) J_z -
\frac{1}{N}J_x^2,  \ee where $\gamma = \frac{\omega_0}{\sigma}  =
\eps t$. In the basis of $\{|j \rangle \equiv
|\frac{N}{2},2j-2-\frac{N}{2} \rangle \}$, the matrix elements
have the form

\bea H_{j,j}&=& 1 - j +\frac{2}{N}(j-1)^2 -  \gamma (-N/2 - 2 + 2j) ,  \\
     H_{j,j+1} &=& -\frac{1}{4 N} \sqrt{2j (2j-1)(N+1-2j)(N+2-2j) }.  \nonumber
 \eea

\begin{figure}
\includegraphics[width=80mm]{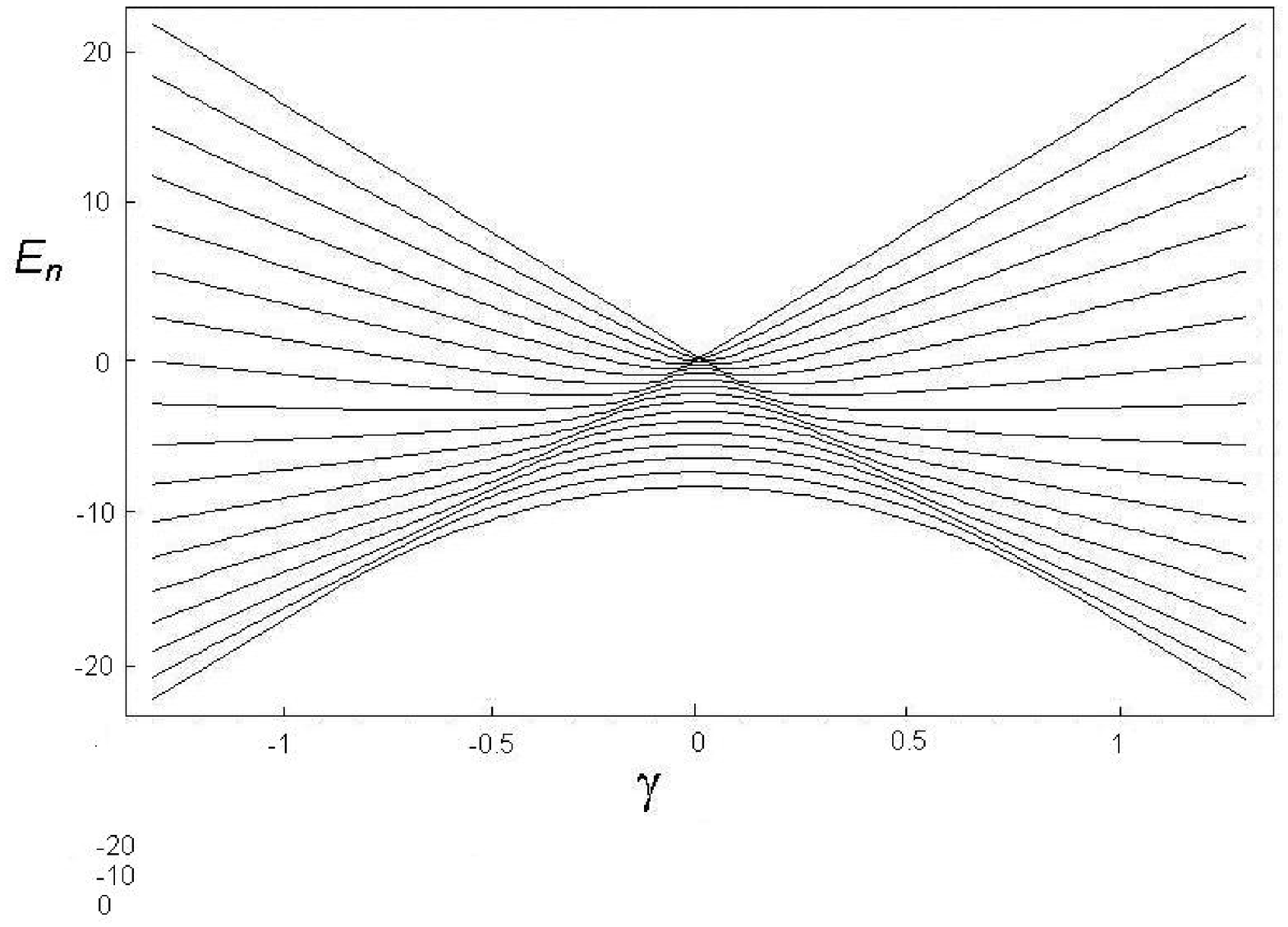}
\includegraphics[width=80mm]{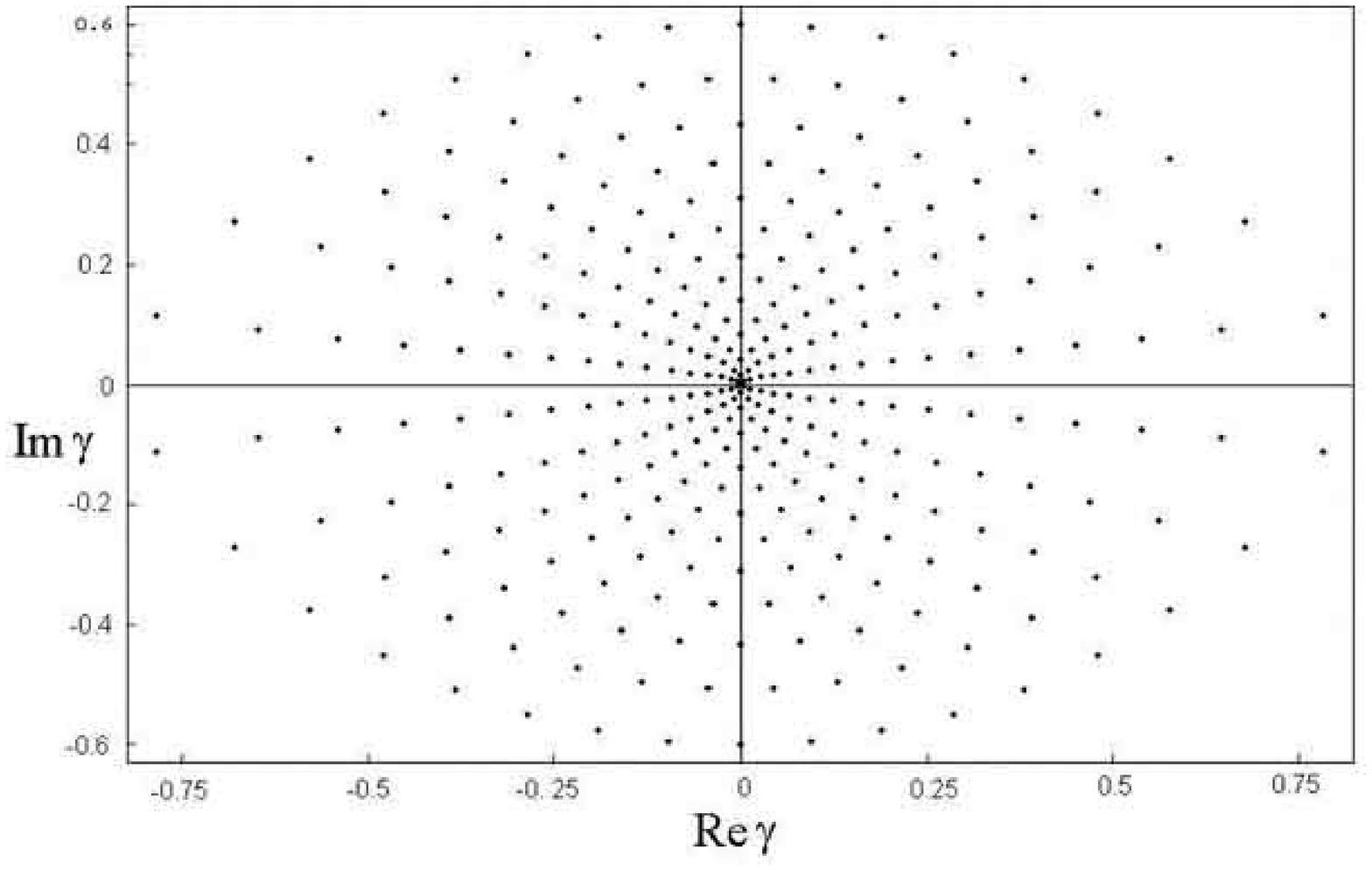}
 \caption{Energy levels of the LMG model $E_n(\gamma$)(upper panel) and its branch points in the plane
 of complex $\gamma$ (bottom panel).} \label{Lipkinlevels}
\end{figure}

The energy levels are depicted on Fig.(\ref{Lipkinlevels})a.
Fig.(\ref{Lipkinlevels})b  shows the locations of the branch
points in the complex $\gamma-$plane. At extremely slow sweeps, in
the lowest order of the sweeping rate the probability of
transition from $i-th$ to $j-th$ eigenstate is determined by the
Dykhne formula \cite{Dykhne} and depends on the integral around
the branch point connecting $i-th$ and $j-th$ eigenvalues
$E_{i,j}(\gamma)$. The regular structure seen in the picture
suggests that there is a universality in the distribution of
probabilities in this (extremely slow) regime, the analysis of
which we postpone for  future research. Here we concentrate on a
faster, semiclassical regime.

\subsection{Classical ensembles and quantum dynamics}

Classical limit of the LMG model is obtained, as above, from the
Heisenberg equations, and is given by

\be H = -\gamma z - \frac{1-z^2}{4}(1+\cos 2\phi), \label{LMGHC}
\ee where $z \in [-1,1]$ is the dimensionless counterpart of
$J_z$, and $\phi \in [-\pi,\pi]$ is the canonically conjugated
phase.

\begin{figure}
\includegraphics[width=80mm]{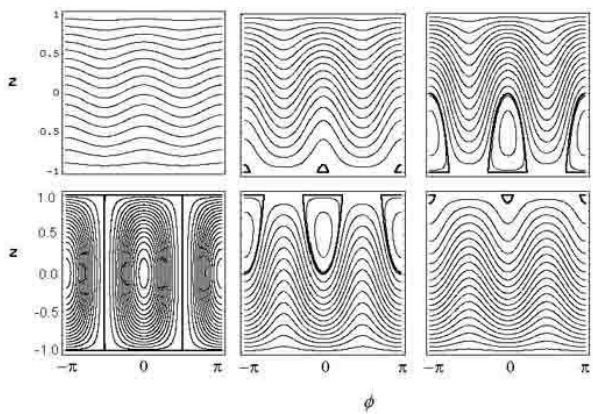}
 \caption{Phase portraits of the LMG Hamiltonian at $\gamma=-3,-0.5, -0.95$ (from left to right in the upper panel) and
$\gamma=0,0.5, 0.95$ (from left to right in the bottom panel).
Note that bifurcations happen at $\gamma = \pm1$. } \label{LMGFP}
\end{figure}

The initial distribution of $z$ following from SU(2) Wigner
function is exponential for large $N$: $P(z) \approx N
\exp[-(1+z)N]$. For large negative $\lambda$, the classical ground
state is at $z=-1$; at $\gamma=-1$ a bifurcation occurs. For
$\gamma \in (-1,1)$ the stable fixed points (s.f.p.) are at
$z=\gamma, \phi=0,\pi$. Phase portraits are shown in
Fig.(\ref{LMGFP}).
It is easy to demonstrate that a passage through bifurcation at
$\gamma=-1$ is similar to that happening in the Hamiltonian system
(\ref{PolkovnikovF}) with $L=0$, studied in Section II. Moreover,
we do not have complications arising from $L \ne 0$ initial
conditions anymore.  Also, a passage through the bifurcation at
$\gamma=1$  happening with increasing $\gamma$ is analogous to the
"inverse" sweep mentioned in Section II. We consider several
scenarios below.


Consider firstly a scenario where  $\gamma(t)$ is changed from a
large negative value  to zero.  The quantum system in the end of
the sweep is described by the Kerr Hamiltonian $-S_x^2$
\cite{Ueda}. In the short-time evolution after the sweep, both
semiclassical and quantum systems exhibit damped oscillations of $
\langle J_z \rangle $. Damping comes from dephasing of the TWA
classical trajectories in the anharmonic effective potential. In
the process of dephasing, $ \langle J_z \rangle$ exhibits decaying
oscillations around its mean (time-averaged) value which we denote
$\overline{\langle J_z \rangle} $.

\begin{figure}
\includegraphics[width=80mm]{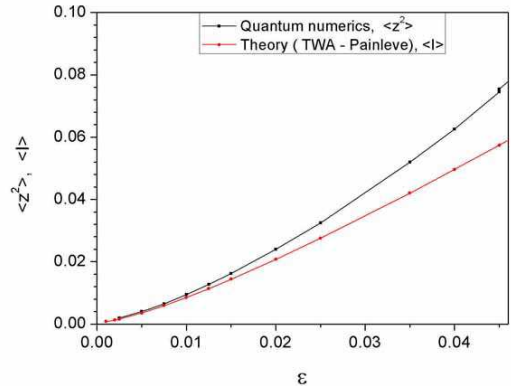}
 \caption{Comparison of the dispersion of $J_z$ in a quantum system
 $\overline{\sigma(J_z)}= (\overline{ \langle J_z^2 \rangle} -
 \overline{\langle J_z \rangle^2 } )/(N/2)^2 \approx \overline{\langle J_z^2 \rangle}/(N/2)^2$
 and the classical prediction from the mean change in classical action of the TWA ensemble,
 $\overline{ \langle I \rangle}$ (Eq. \ref{dispersionprediction}).} \label{Fdispersion}
\end{figure}

As a signature of nonadiabaticity, let us consider the
time-averaged dispersion of $\langle J_z \rangle (t)$, or the mean
value of the second moment $\overline{\langle J_z^2 \rangle} $
around which the dispersion of $\langle J_z \rangle (t)$
oscillates. In the classical system, at the final value
$\gamma=0$, the Hamiltonian linearized around its s.f.p. $\{
z=0,\phi=0\}$ is $H = \frac{z^2}{2}+\frac{\phi^2}{2}$. Therefore
the dispersion of $z$ is related to the mean value of the
classical action in the first approximation. That is, assuming a
uniform distribution of the final phase of trajectories of the
classical wavepacket (i.e., after dephasing), one obtains

\bea \overline{\langle z^2 \rangle} &=& \overline{\langle I \rangle} \nonumber \\
\approx \int_0^{\infty} \Bigl[ \frac{3(1
+z_0)}{4}&-&\frac{\eps}{\pi} \ln \left(
\frac{\pi (1+z_0)}{2\eps}\right) \Bigr] N e^{-(1+z_0)N} dz_0 \nonumber \\
= \frac{3}{4N} &+& \frac{\eps}{\pi} \left( C_{\gamma} + \ln
\left(\frac{2\eps N}{\pi} \right)\right) \nonumber\\ &\approx&
\frac{\eps}{\pi} \left[ C_{\gamma} + \ln \left(\frac{2\eps N}{\pi}
\right)\right],   \label{dispersionprediction} \eea where
$C_{\gamma} \approx 0.577$ is the Euler constant. Here, $\langle
.. \rangle$ denotes averaging over the ensemble, while overline
denotes time-averaging. Dispersion of $\langle J_z \rangle$ in a
quantum system with $N=1024$ is compared with
Eq.(\ref{dispersionprediction}) in Fig. \ref{Fdispersion}. One can
see a remarkable coincidence of the theoretical predictions and
numerics at slow sweeps.  Deviations at fast sweeps are due to
inapplicability of the Painleve mapping. We note a universality of
the obtained formula: the equivalent linear plus
linear-logarithmic law was derived in Section II.


Let us consider the semiclassical distribution in action space
after the sweep. In Section II we already derived the distribution
$\bar P(I)$ coming from the phase-dependent part of change in the
action. We consider now in detail how it is summed up to the full
distribution.


\begin{figure}
\includegraphics[width=80mm]{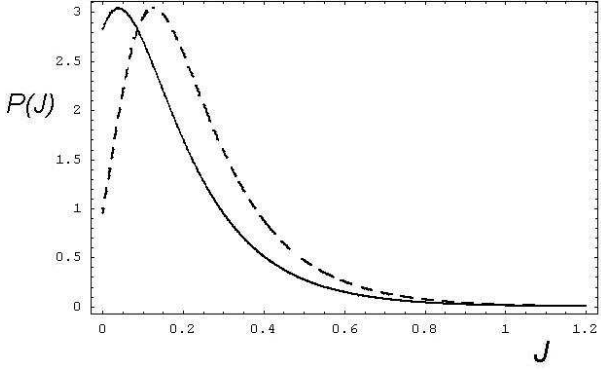}
 \caption{The distribution ${\ cal P}(\tilde J)$ in the rescaled action space after the sweep, Eq. (\ref{Distribution}). Solid line: $\eps=0.015$.
 Dashed line: $\eps=0.005$.}
\label{DistributionP}
\end{figure}

\begin{figure}
\includegraphics[width=80mm]{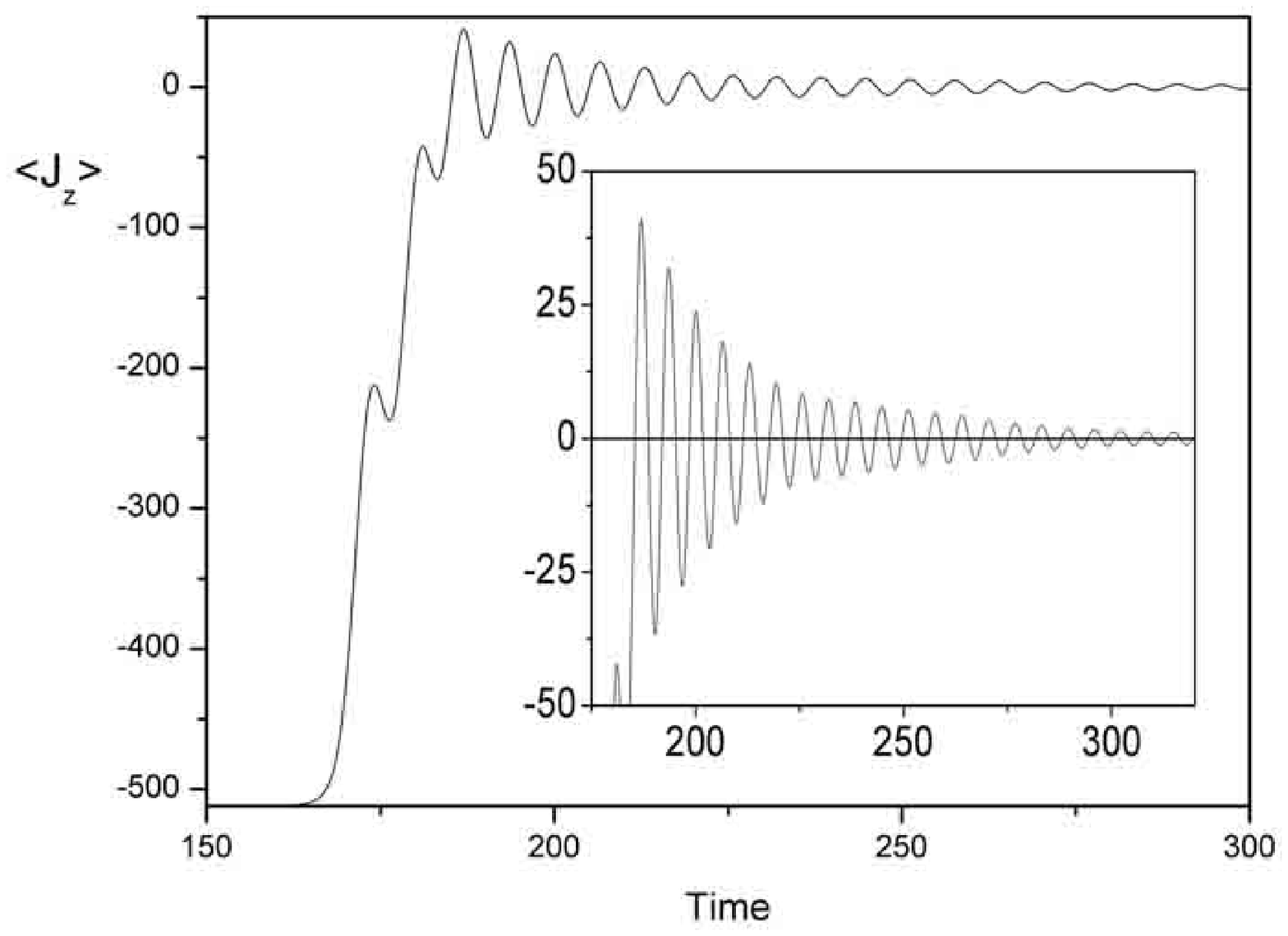}
\includegraphics[width=80mm]{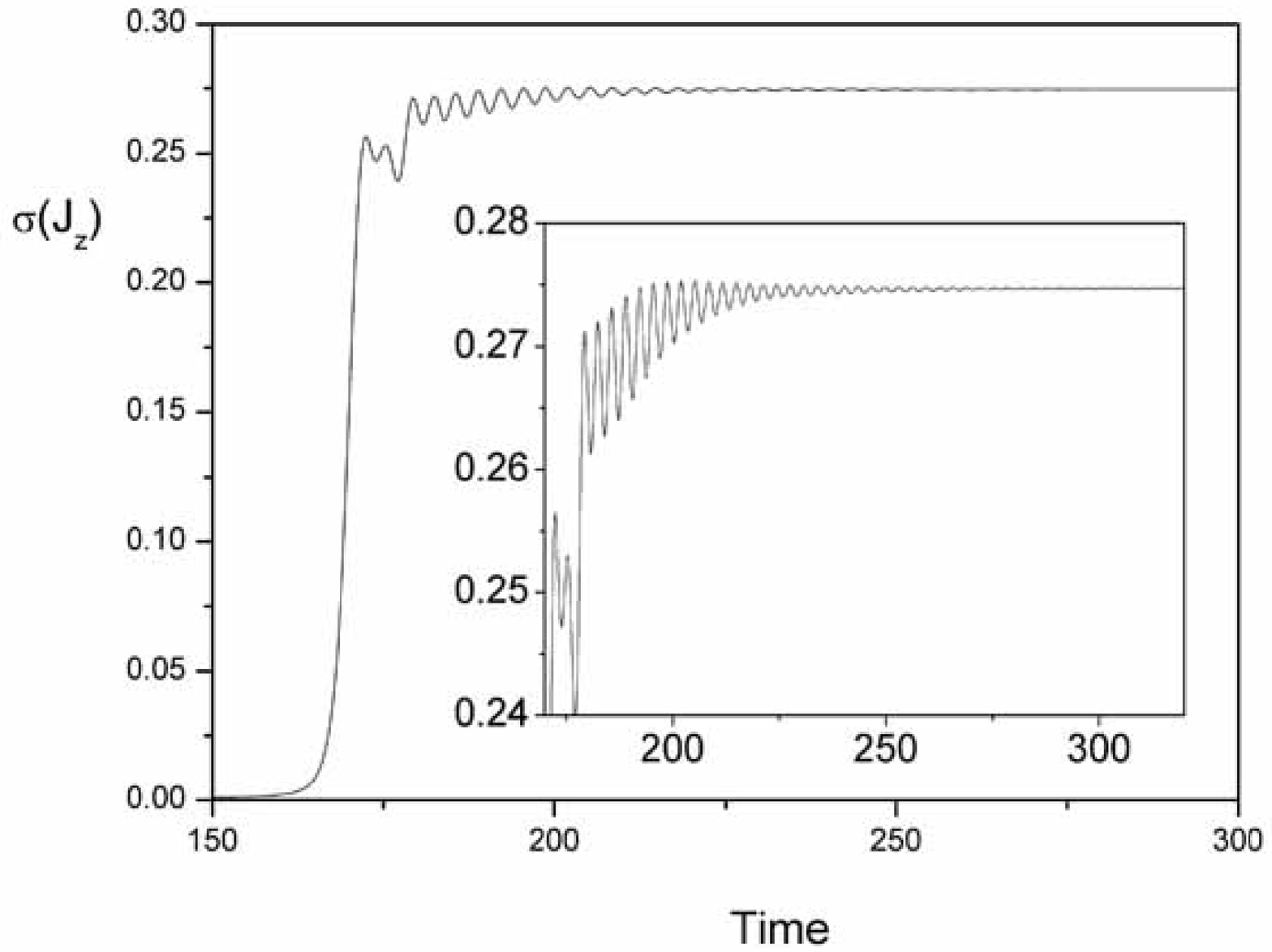}
 \caption{Dynamics of $J_z$ (upper panel) and the dispersion of $J_z$ (bottom panel) in
the quantum system, N=1024,  $\eps=0.045$.} \label{Jzdynamics}
\end{figure}

\begin{figure}
\includegraphics[width=80mm]{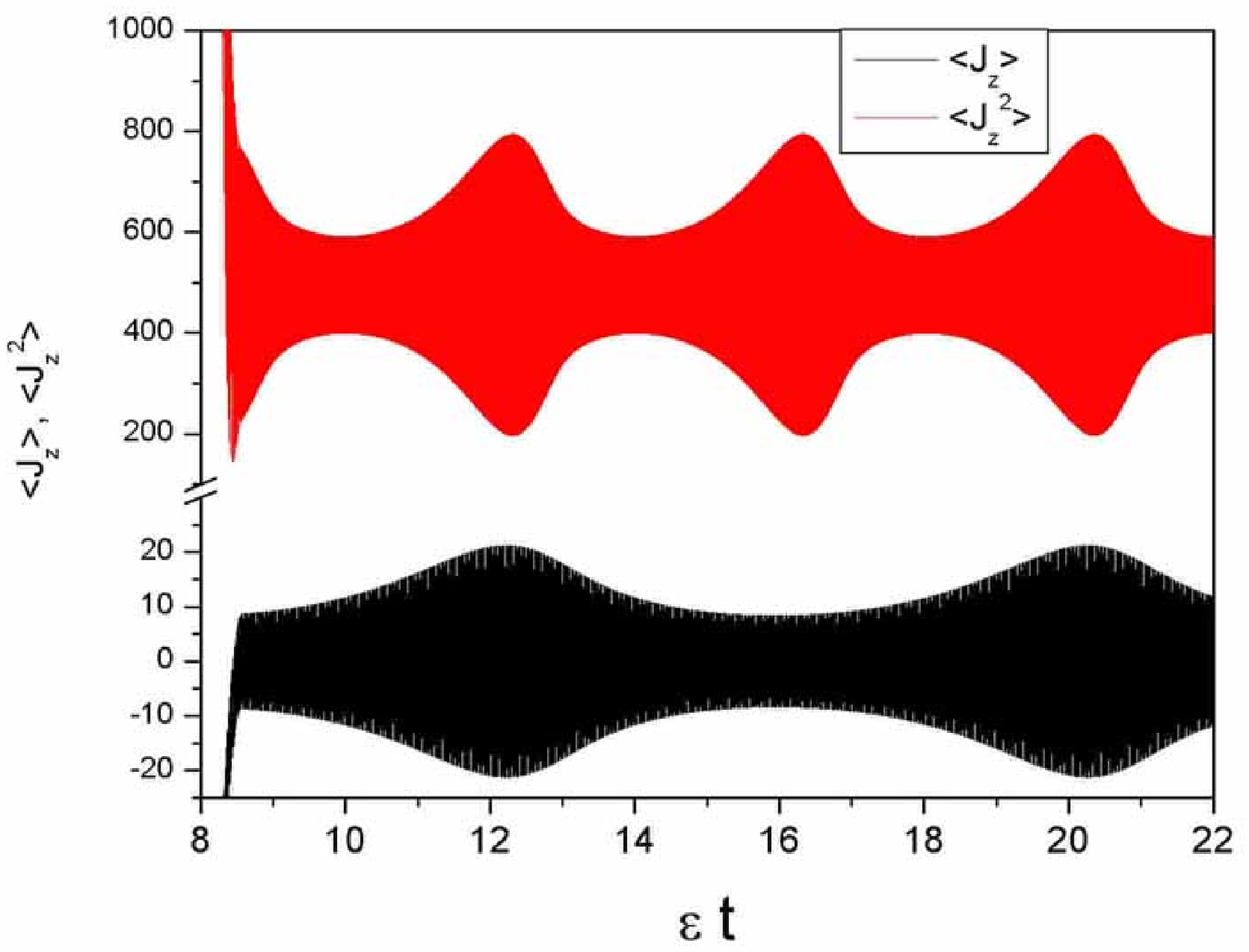}
\includegraphics[width=80mm]{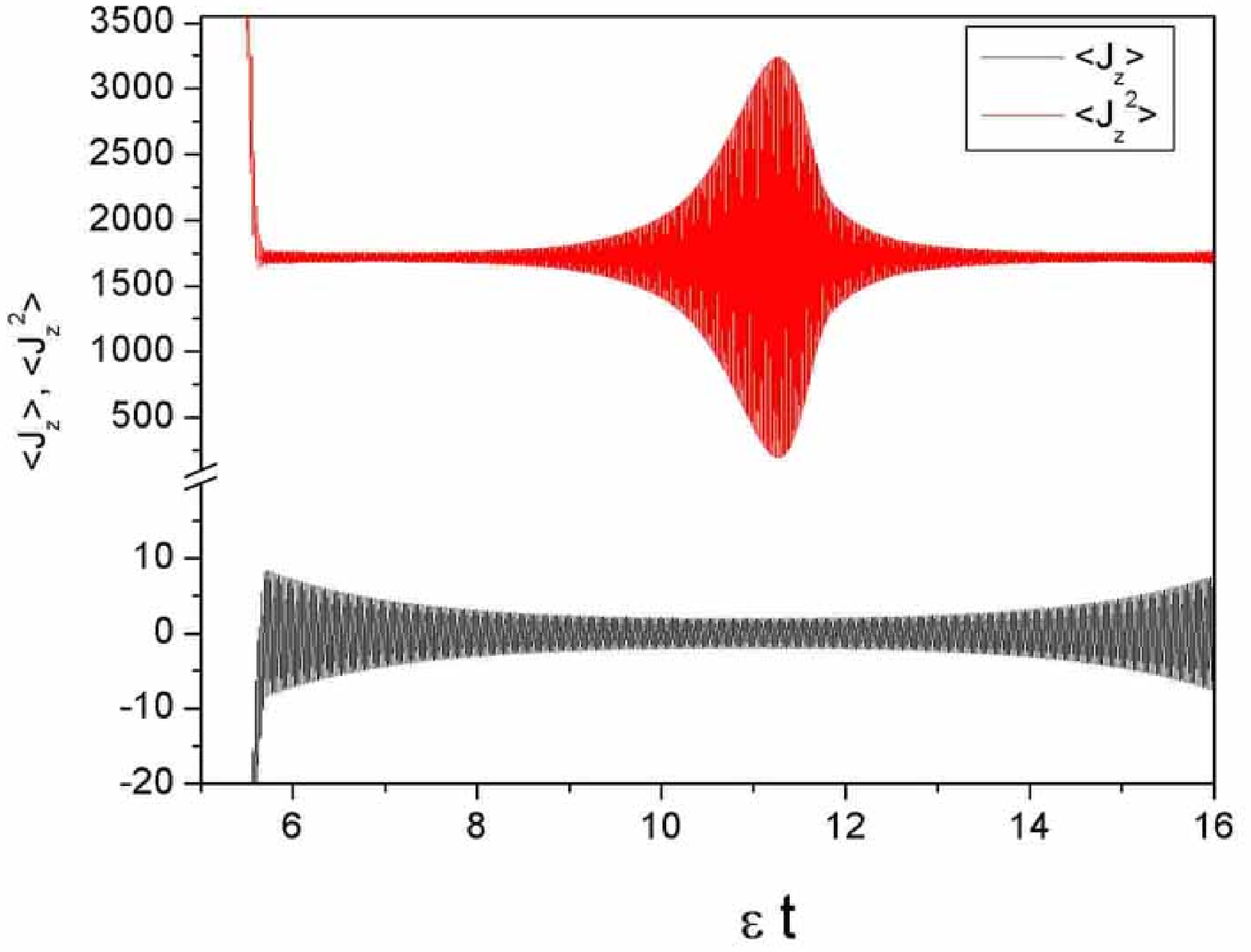}
\includegraphics[width=80mm]{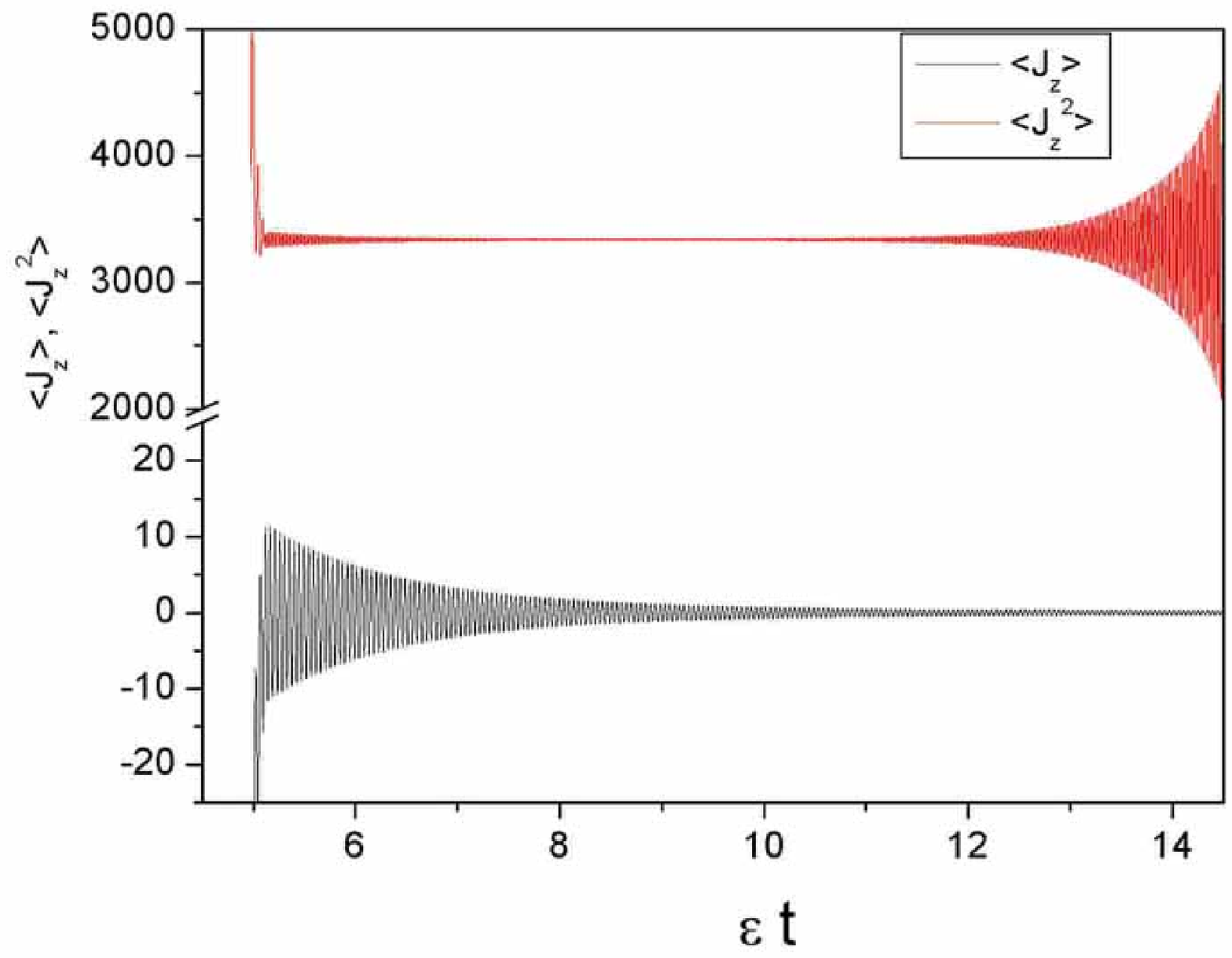}
 \caption{Dynamics of $ \langle J_z \rangle , \langle J_z^2 \rangle$ for $\eps=0.0025, 0.005, 0075$ (from top to bottom)
in the quantum system, N=1024. Note  the different scales in each
figure for  $\langle J_z \rangle$ and $ \langle J_z^2 \rangle$.
The time scales in both figures look shifted because we use smooth
turn-on and turn-off of sweeping of the parameters. Both
magnitudes after the sweep undergo oscillations around its
time-averaged values, $\overline{ \langle J_z \rangle}$ and
$\overline{ \langle J_z^2 \rangle}$, correspondingly (overline
denotes time-averaging here, while $\langle .. \rangle$  denotes
instanteneous mean value of an operator, i.e. averaging over the
state). It can be seen that $\overline{ \langle J_z \rangle}=0$,
while $\overline{ \langle J_z^2 \rangle}$ is a function of the
sweeping rate $\eps$, as shown in Fig. (\ref{Fdispersion}). }
\label{Jzdynamics2}
\end{figure}


Let us work with the rescaled action $ \tilde I = \frac{I}{4
\eps}$, where $I$ is the action of the Hamiltonian (\ref{LMGHC}).

Following the analysis of Section II, action after the sweep
through the critical $\gamma=-1$ ($\tilde I_+$) is related to the
initial rescaled action $\tilde I_-$ as

\be \tilde I_+ = \tilde I_- - \frac{1}{2\pi} \ln \sqrt{\exp (2\pi
\tilde I_-  ) - 1 } - \frac{1}{2\pi} \ln 2 \sin \pi \xi' . \ee

Let us consider firstly the phase-dependent part of change in the
action $-\frac{1}{2\pi}\ln 2\sin \pi \xi$. It leads to the
distribution \be \bar{\cal  P}(\tilde I_f)=\frac{2}{ \sqrt{
4\exp(4\pi \tilde I_f)-1}} \ee in the final action space.
Additionally, each slice ${\cal{A}}_{{\tilde I}_-}$ gets a
different phase-independent change  \be   \tilde I_+ (\tilde I_-)
= \tilde I_- - \frac{1}{2\pi} \ln \sqrt{\exp (2\pi \tilde I_-  ) -
1 } \approx \frac{3}{4} \tilde I_- - \frac{1}{4\pi} \ln \ 2\pi
\tilde I_- . \label{phaseindependent} \ee The final distribution
is \be {\cal P}(\tilde I) = \int \bar{\cal P}(\tilde I - \tilde
I_+ (x))W(x) dx, \ee i.e., it is an integral of the distribution
$\bar {\cal P}(\tilde I - \tilde I_+(\tilde I_-) )$ originating
from a phase-dependent change in action $\bar {\cal P}(\tilde I)$
of an ensemble $ {\cal A}_{\tilde I_-}$ shifted by value of
phase-independent change in action of that ensemble over the
initial exponential distribution of initial actions $W(\tilde I_-)
= N \exp[-\tilde I_- N]$. After simple manipulations, neglecting
the term $\frac{3}{4} \tilde I_-$ in (\ref{phaseindependent}), we
get

\be {\cal P}(\tilde I) = \sqrt{ 8N\eps} \exp(-2 \pi \tilde I)
\exp(-\frac{\eps N}{2\pi} \exp[-4 \pi \tilde I]) ,
\label{Distribution} \ee see Fig. \ref{DistributionP}.

 It is easy to see that normalization of the
distribution (\ref{Distribution}) is less than 1: the total
probability is $\int P(I)dI = \mbox{Erf}{\sqrt{\frac{\eps
N}{2\pi}}}$, which is close to $1$ only for $\eps N \gg 1$. This
discrepancy is due to the approximations in
(\ref{phaseindependent}), which were necessary to obtain the
analytical expression (\ref{Distribution}).  Nevertheless, even at
$\eps N \sim 1$ the obtained analytical formula correctly
describes distributions at small final actions $I_f \ll 1$. We
believe that $P(I_f)$ may be fitted to the Gumbel distribution as
done in \cite{Polkovnikov}.

The distribution is manifested, in particular, in the final
dispersion of $\langle J_z \rangle$. Indeed, for a Morse
oscillator $H=-\frac{1}{2}(1-I)^2$ a trajectory with $I= $const
has $\bar{z^2}(t) = \frac{1-(1-I)^2}{2}=(I -\frac{I^2}{2})$. The
dispersion of $z$ of the whole wavepacket is therefore \be
\bar{z^2} = \int (I-\frac{I^2}{2}) {\cal P}(I)dI =
\bar{I}-\frac{\bar{I^2}}{2}. \ee This formula shows that
anharmonicity of the final Hamiltonian provides an additional
means of measuring the final phase-space distribution.  Indeed,
now the dispersion of $z$ depends not only on the mean value of
$I$, but also on its second moment. The additional contribution
from anharmonicity is given by \bea \delta \bar{z^2}
&=&-\frac{2}{3}\frac{\eps^2}{\pi^2} \Bigl( \pi^2 \nonumber\\ &+& 6
(C_{\gamma}+\ln N)^2  + 6 \ln \left(\frac{2 \pi}{\eps N^2}-2
C_{\gamma} \right) \Bigr) \ln \left( \frac{2\pi}{\eps} \right).
\nonumber \eea

Having obtained the distribution  ($\ref{Distribution}$) resulting
from "forward" sweep through the QPT, it is interesting to find
out what happens with this distribution if $\gamma$ is slowly
sweeped back, and the system is pushed through the QPT in
"backward" direction. Alternatively,  one may choose not to stop a
sweeping of $\gamma$ at $\gamma=0$, but continue to change it
(linearly in time) beyond the next critical point $\gamma=1$.

So, we consider the second possible scenario which involves a
sweep from $\gamma=-\infty$ to $\gamma = + \infty$.  Then, the
system undergoes a passage through a bifurcation twice: at
$\gamma=-1$ and at $\gamma=1$. The latter passage has a very
different behaviour: in particular, the jump in the classical
action of a trajectory is phase-independent for small values of
initial actions, like in the backward sweep of the model of
Section II.

The adiabatic invariant after the inverse sweep $\tilde I_{++}$ is
given by the following formula \cite{Mitya}:

\be \tilde I_{++} = \frac{1}{2\pi} \ln (1+|p|^2), \ee where

\bea p = p_r &+& p_i,  \quad p_r = \mbox{sign} (\sin c)
\sqrt{(1-b^2)(a^2-1)}, \nonumber\\ p_i &=& a - b\sqrt{a^2-1},
\quad b = \cos c,  \\ c &=& \phi + \tilde f(I_+), \quad a =\exp(2
\pi I_+ ), \eea where $\tilde f(I_+)$ is a function of $I_+$;
$I_+$ denotes the value of the adiabatic invariant after a direct
sweep and before the inverse sweep, say at $\gamma=0$. We assume
that dephasing of phase points happening between the first and the
second passage through bifurcations lead to uniform distribution
of the magnitude of $c$. Denote as $\langle I_{++} \rangle_{c} $
the mean value of $I_{++}$ averaged over the initial phase $c$.
Since

\be I_{++} = \frac{1}{2 \pi} \ln (2a^2 - 2ab \sqrt{a^2-1}),  \ee
we get after averaging over $c$:

\be \langle I_{++} \rangle_c = \frac{1}{2\pi} \ln(a^2+a).  \ee

The average over the whole ensemble is achieved now by averaging
over the distribution of actions ${\cal P}(I_+)$:

\be \langle I_{++} \rangle = \frac{1}{2 \pi} \int dI_{+}
\ln(a^2+a){\cal P}(I_+). \label{meanplusplus} \ee Therefore, the
distribution ${\cal P}(I_+)$ {\em before} the second bifurcation
passage leads to the following prediction for the final values of
$\overline{ \langle J_z \rangle} $:

\bea \overline{\langle J_z \rangle} &=& 2 \eps N \int_0^\infty dI
\sqrt{8  \eps N}\exp[-2\pi I] \times \label{meanJzplusplus}  \\ &
\times& \exp \Bigl[- \eps N \frac{\exp[-2\pi I]}{2\pi} \Bigr]
\frac{\ln(\exp(2\pi I) +\exp(4\pi I))}{2\pi}. \nonumber \eea

The correspondence between theoretical prediction, quantum
numerics and TWA numerics is shown in Fig.\ref{M8P8e}.

We note that up to $\eps \geq  0.1$ TWA numerics closely reproduce
quantum numerics, but starts to considerably deviate from TWA
theoretical prediction at $\eps \sim  0.1$. the assumption of
uniform phase distribution before the second bifurcation passage
seems to be not fulfilled here. Also, at $\eps \leq \frac{1}{N}$
deviations between TWA and quantum numerics can be seen which can
be attributed to finite-size corrections and interference effects.


\begin{figure}
\includegraphics[width=80mm]{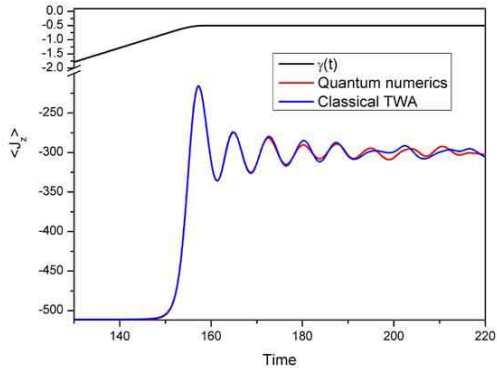}
 \caption{Sweeping from $\gamma=-8$ to $\gamma=-0.5$. The system undergoes a phase transition at $\gamma=-1$
and the sweeping is stopped at $\gamma=-0.5$. Short-time evolution
exhibits remarkable coincidence of quantum and semiclassical
dynamics. $\eps = 0.05, N=1024$.
} \label{M8M05}
\end{figure}

\begin{figure}

\includegraphics[width=80mm]{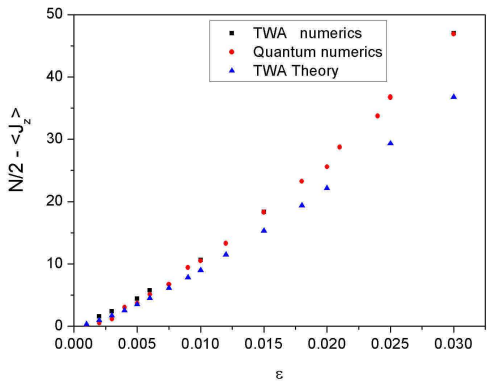}
 \caption{Sweeping from $\gamma=-8$ to $\gamma=16$. The system undergoes two consequential
phase transitions at $\gamma= \mp 1$. The final values of $
\overline{\langle J_z \rangle}$ are shown, extracted from quantum
numerics, classical TWA numerics, and TWA-Painleve theoretical
prediction. Theoretical TWA prediction (Eq.(\ref{meanJzplusplus}))
is depicted as triangles; circles: quantum numerics, and TWA
numerics is shown by squares. We note that up to $\eps \geq  0.1$
TWA numerics closely reproduce quantum numerics, however slightly
deviate from TWA theoretical prediction (\ref{meanJzplusplus}).
Also, at very slow sweeps $\eps \leq \frac{1}{N}$ deviations
between TWA  and quantum numerics can be seen.
} \label{M8P8e}
\end{figure}



\section{Dicke Model with counterrotating terms}

\subsection{The model}

For many applications, counterrotating terms should be included in
the model Hamiltonian:

\be H = \omega_0 J_z + \bar \omega b^\dagger b + \frac{
\bar\lambda}{\sqrt{2J}} (b^\dagger + b)(J_+ + J_-), \label{Dicke2}
\ee here the angular momentum operators describe an ensemble of
two-level systems of level-splitting $\omega_0$ in terms of a
collective spin of length $J=N/2$. The field mode frequency is
$\bar \omega$. A physical realization is e.g. an ensemble of
two-level atoms in a cavity; ground state of such a system
corresponds to the absence of photons (field vacuum) and complete
angular momentum inversion ($J_z=-N/2$, all atoms in the lowest
state).

 We introduce a rescaled coupling $\lambda = \tilde
\lambda/\omega_0$, a dimensionless frequency
$\omega=\bar\omega/\omega_0$, and a dimensionless time $t' =
\omega_0 t$. The Hamiltonian now depends on two parameters \be H =
J_z + \omega b^\dagger b + \frac{\lambda}{\sqrt{N}} (b^\dagger +
b)(J_+ + J_-). \ee There are several different approaches to
obtain semiclassical equations of motion:  through coherent states
\cite{Furuya}, Holstein-Primakoff transformation \cite{Brandes},
and
 from Heisenberg equations of motion through straightforward c-number formalism \cite{KPrants}.
We choose to use the latter approach, but with a very important
additional detail: we combine it with the Wigner function approach
of Altland et al. \cite{Polkovnikov}. That is, we firstly
investigate classical equations of motion analogous to that of
\cite{KPrants}, for the case of sweeping the coupling through the
quantum phase transition. Then, we prepare an ensemble of
classical trajectories corresponding to our initial quantum state
(field vacuum and complete angular momentum inversion,
$J_z=-N/2$), and compare quantum and classical dynamics. The
magnitude of any quantum observable is obtained by averaging over
the classical ensemble.

\begin{figure}
\includegraphics[width=75mm]{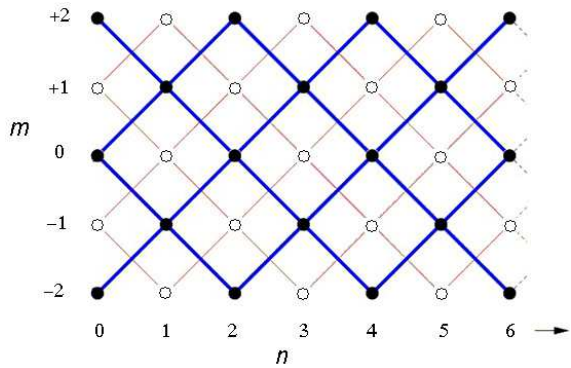}
 \caption{ The Hilbert space of the Dicke model, for $2j=N=2$. $m$ is the z-projection of angular momentum,
 $J_z$; $n$ is  number of photons;  see also \cite{Brandes}. }
\end{figure}

\subsection{Classical dynamics}

Similar to the case of Eq.(\ref{DickeCl}) in Section II, classical
dynamics is obtained from Heisenberg equations of motion and is
descibed by five dynamical variables:

\bea
\dot{x} &=& -y \nonumber\\
\dot{y} &=& x - 2 \lambda e z  \nonumber\\
\dot{z} &=& 2 \lambda e y \label{DickeCl2}\\
\dot{e} &=& \omega p \nonumber\\
\dot{p} &=& -\omega e - 2 \lambda x. \nonumber \eea
\begin{figure*}
\includegraphics[width=80mm]{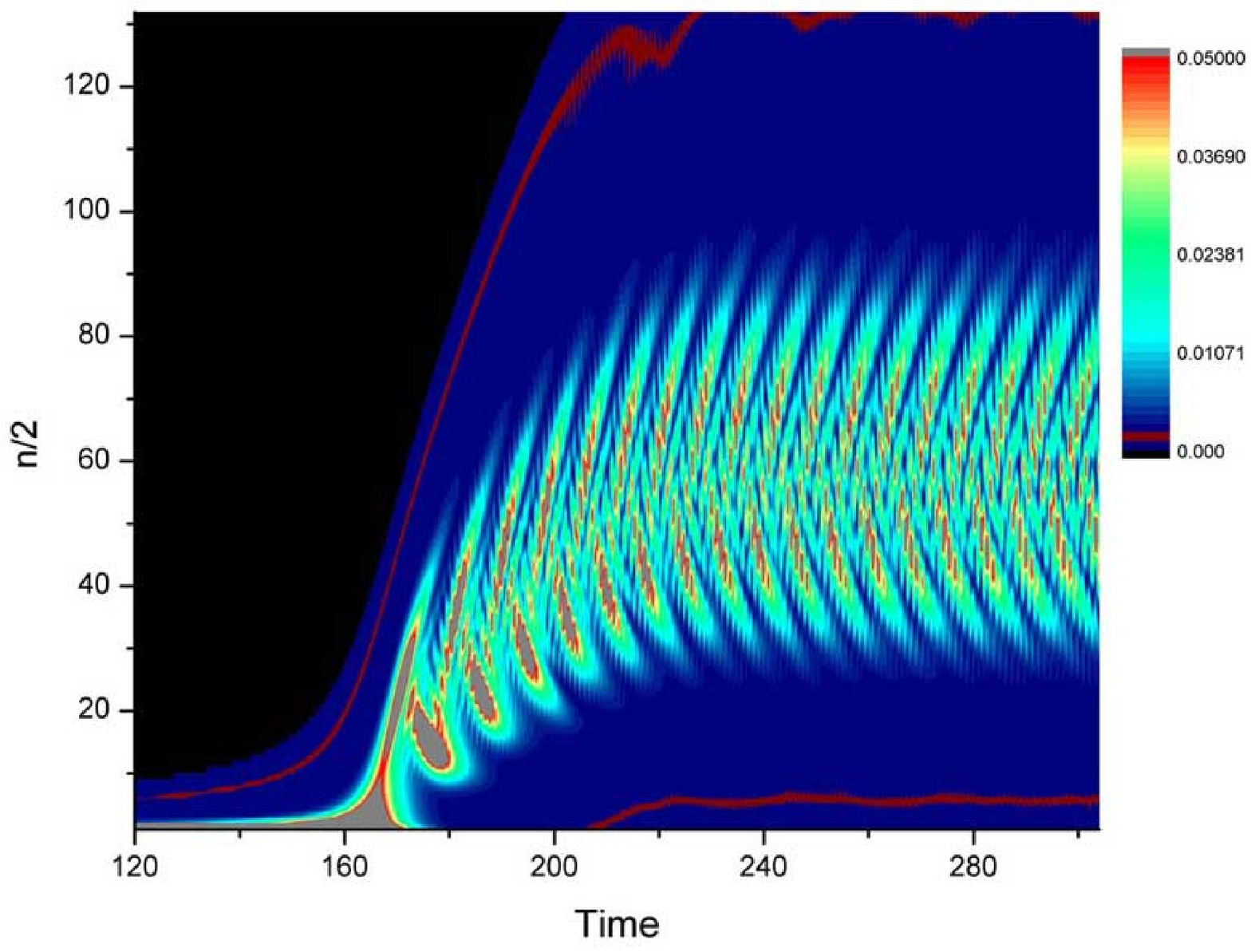}
\includegraphics[width=80mm]{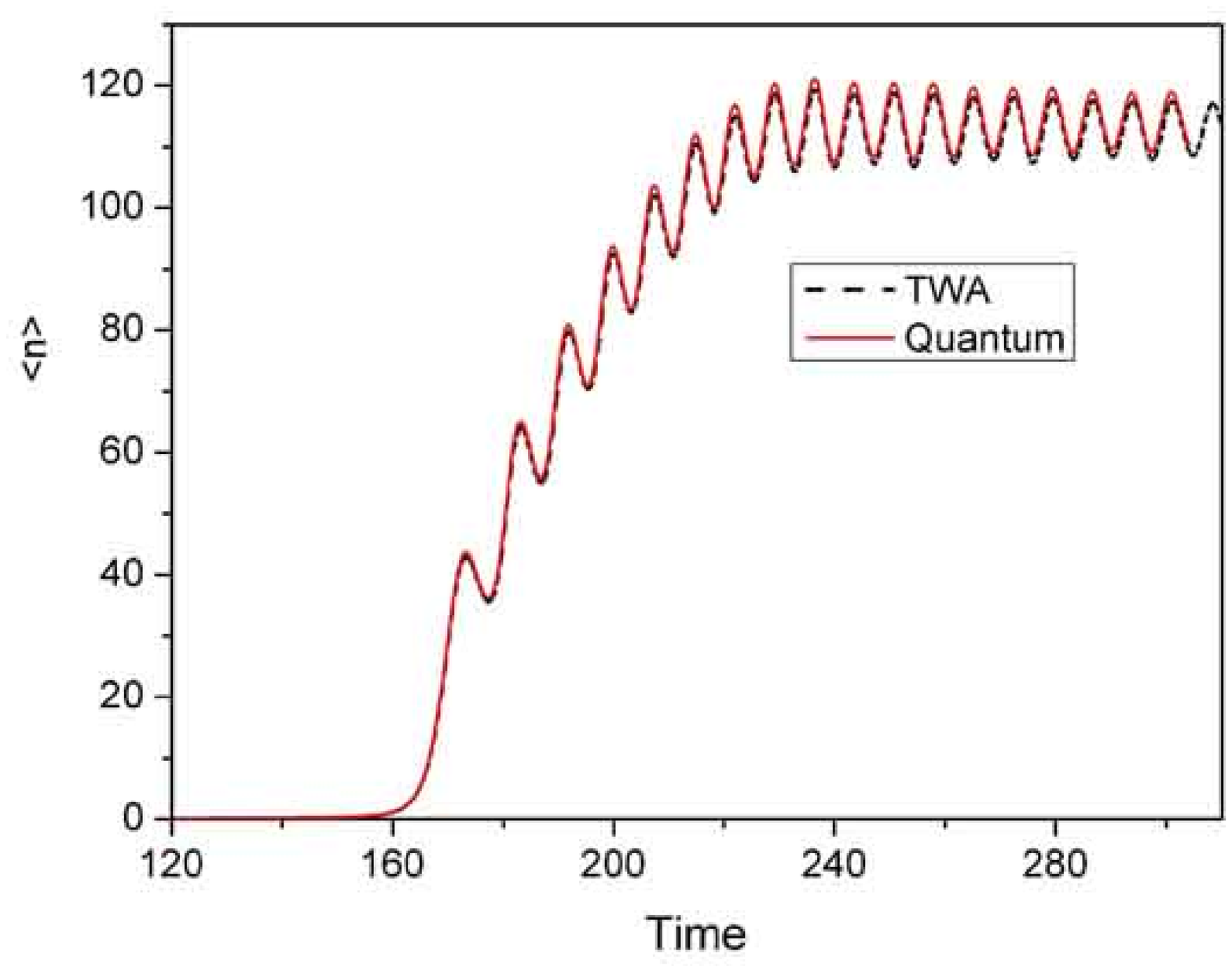}
 \caption{Time evolution of the photon distribution in quantum numerics (left), and the mean photon
  number in quantum and TWA numerics (right) for a sweep
 from $\lambda=0.25$ to $\lambda=0.75$ in the Dicke model on resonance
$\omega=1$. The parameters are $\eps=0.01, N=250$. The
correspondence between semiclassical TWA and quantum calculations
is nearly perfect.  }
\includegraphics[width=80mm]{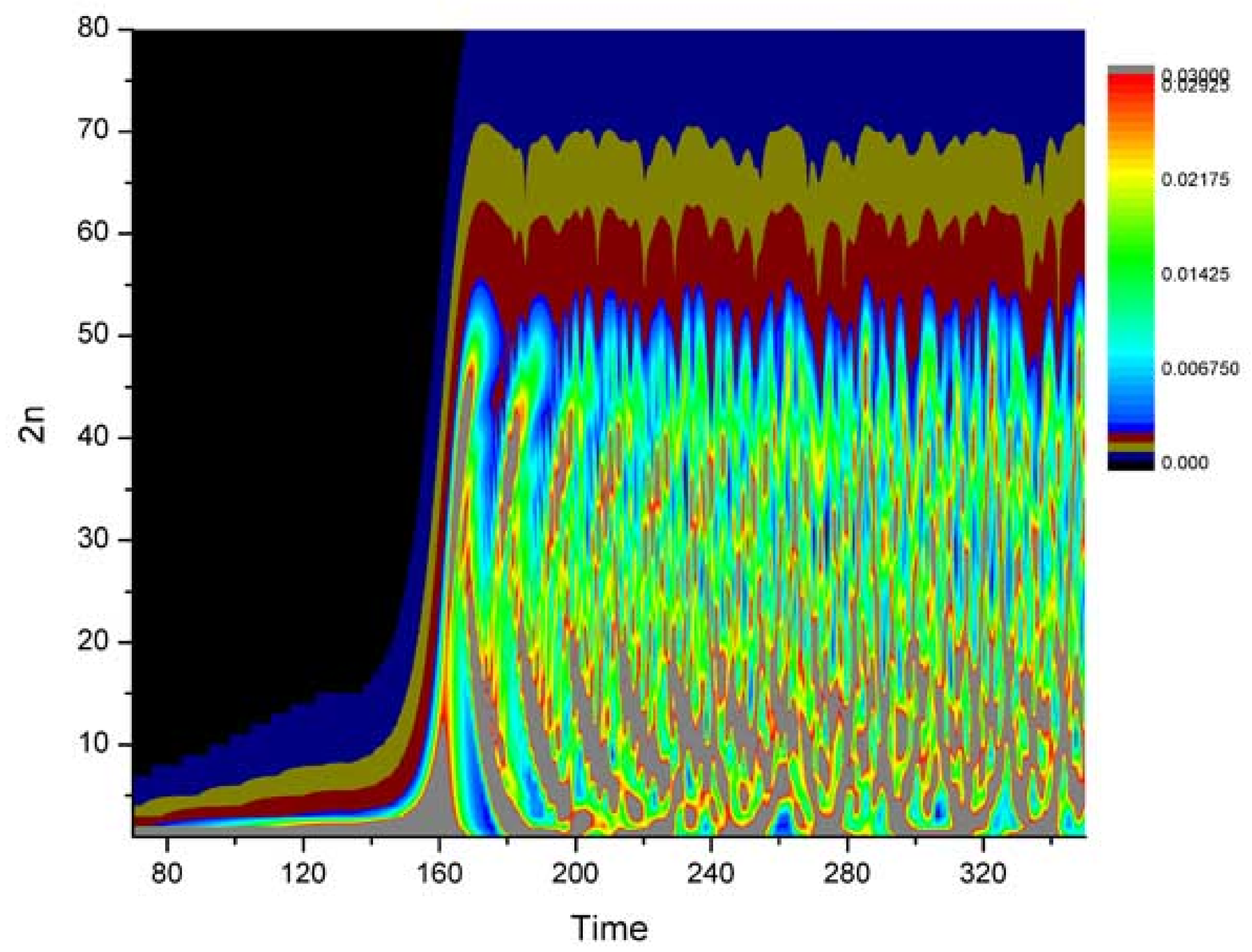}
\includegraphics[width=80mm]{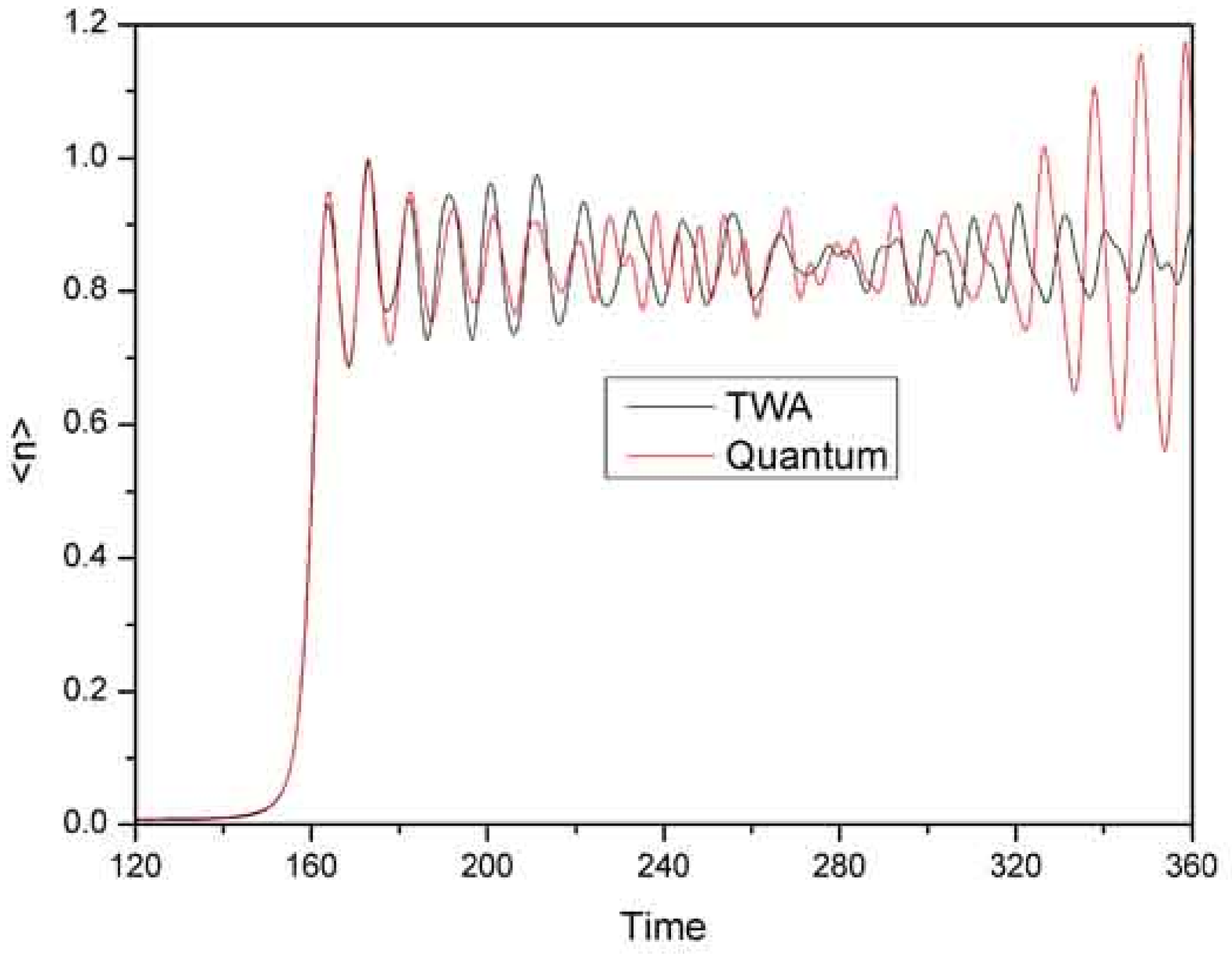}
 \caption{Time evolution of the photon distribution in quantum numerics (left), and the mean photon
  number in quantum and TWA numerics (right) for a sweep
 from $\lambda=0.45$ to $\lambda=0.55$ in the Dicke model with large difference of frequencies
$\omega=50$. The parameters are $N=500, \eps=0.04$.  The
correspondence between  semiclassical TWA and quantum calculations
degrades with time.} \end{figure*}

\begin{figure*}
\includegraphics[width=80mm]{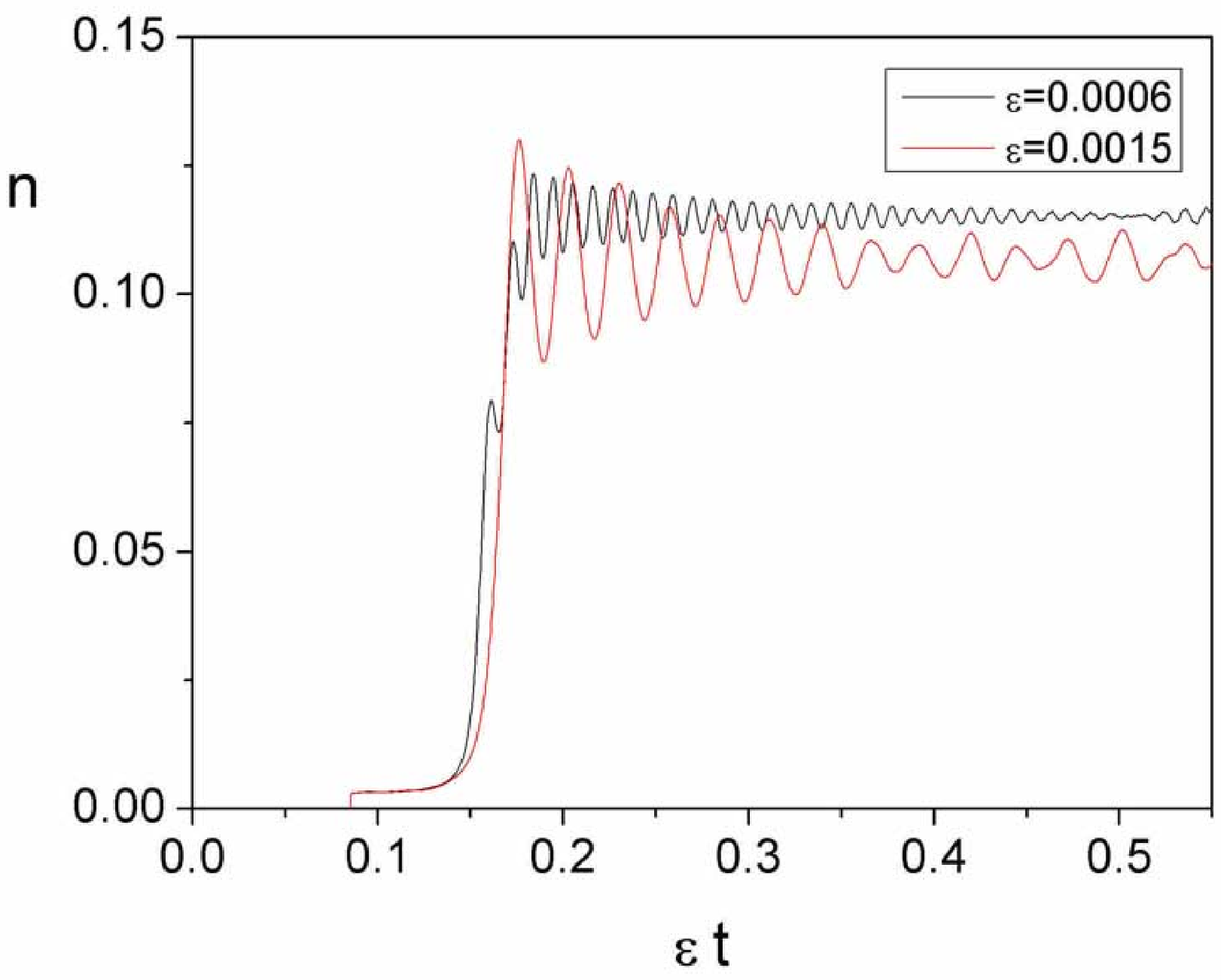}
\includegraphics[width=80mm]{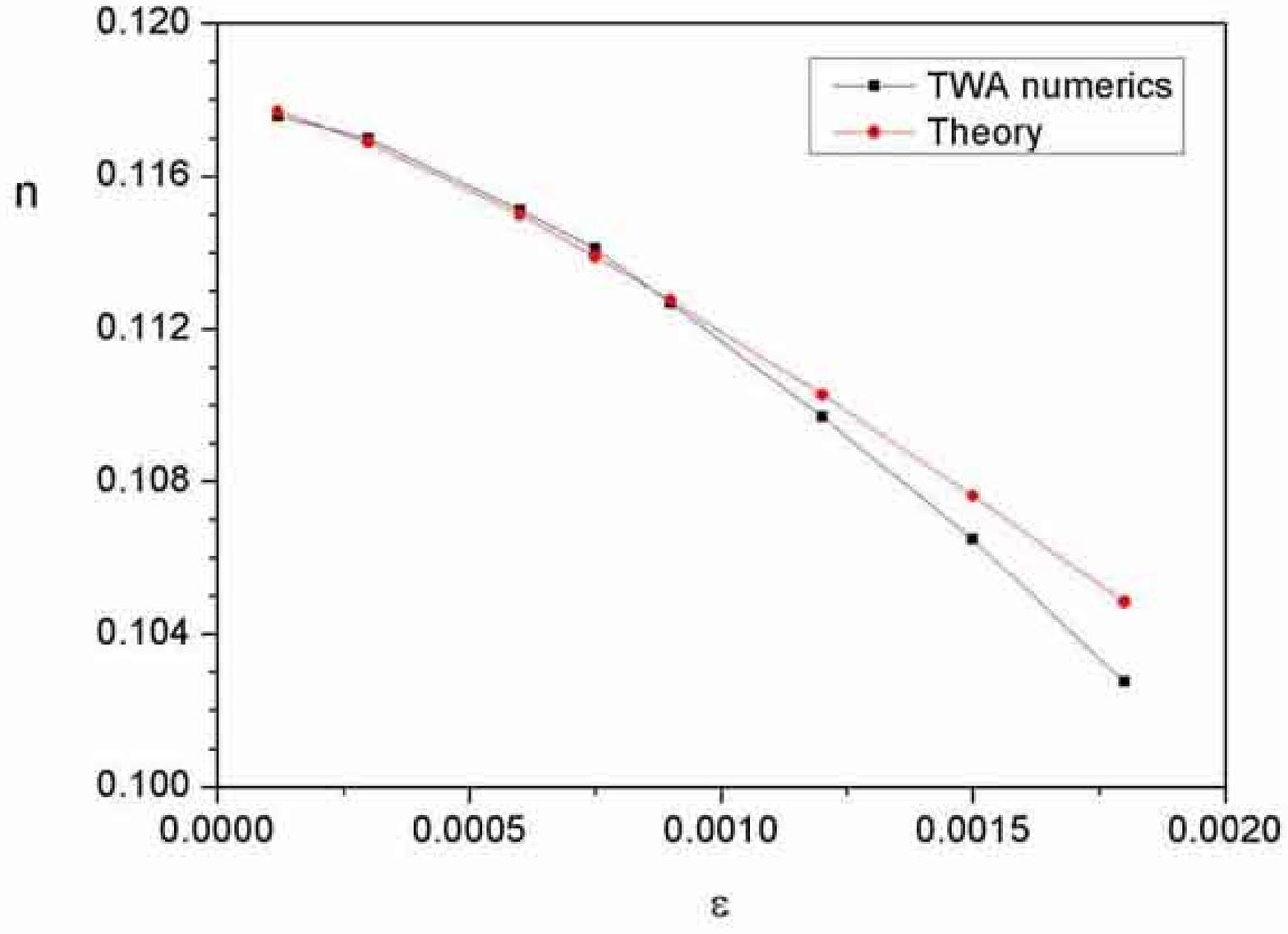}
\includegraphics[width=110mm]{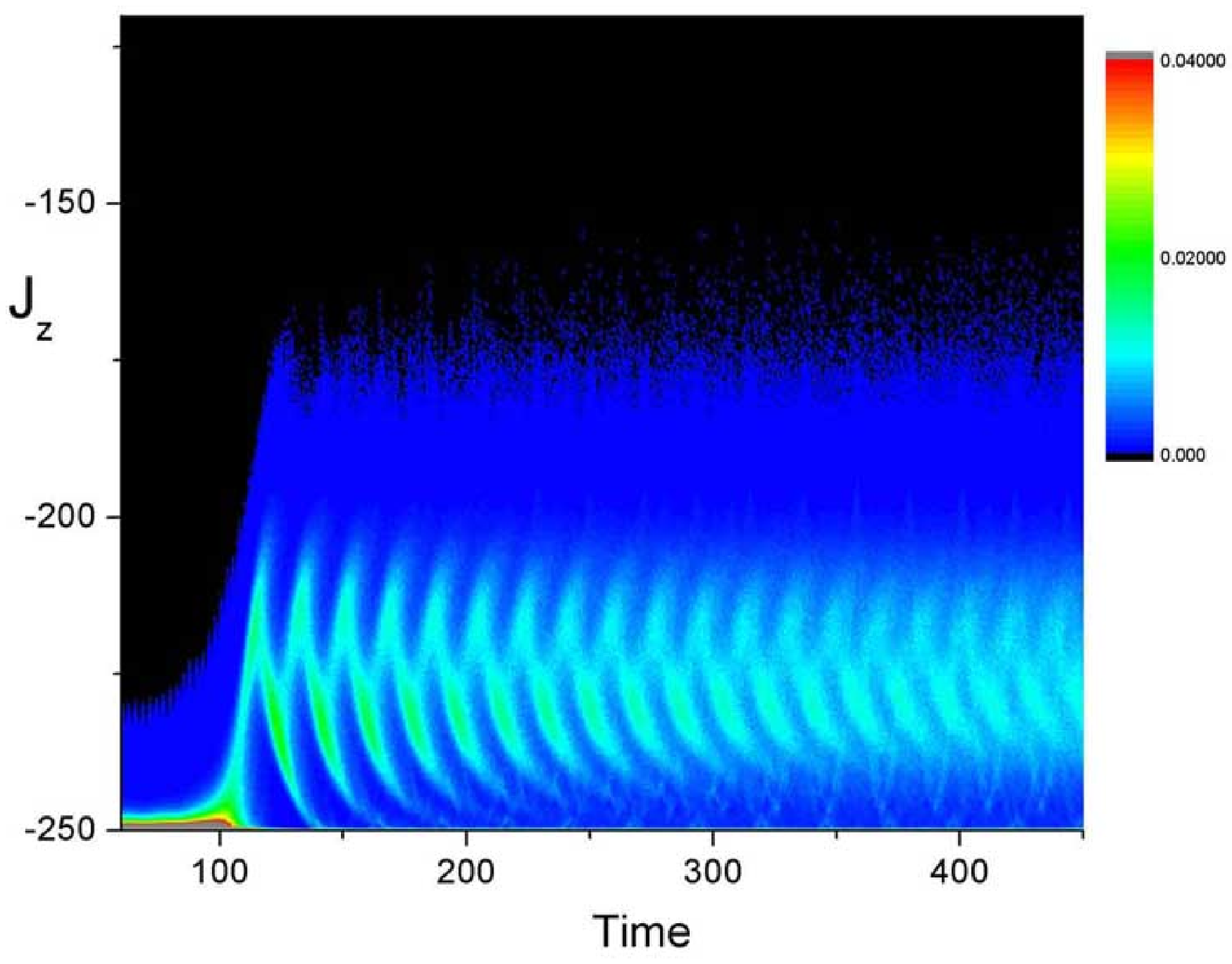}
 \caption{Sweep through the critical coupling from $\lambda=0.47$ to $\lambda_f=0.53$,
$\omega=1$, $N=500$. TWA numerics and theoretical TWA-Painleve
prediction is given. Top left: time evolution of the photon number
for different sweeping rates. The photon number undergoes damped
(due to semiclassical dephasing) oscillations around a mean value.
Faster sweeps lead to shift of the mean value downwards and
increase of the initial amplitude of these oscillations. Top
right: the mean value (i.e., time-averaged value) of the photon
number after the sweep for different sweeping rates, comparison of
the TWA numerics and the theoretical prediction
(Eqs.(\ref{nbar2},\ref{I2final2})). Bottom: dynamics of
distribution of $J_z$ during the sweep in TWA numerics (similar
dynamics for the photon distribution).}
\end{figure*}

 The integrals of motion are
$r=x^2+y^2+z^2=1$ (irrespective of the time-dependence of the
parameter $\lambda$) and  $\frac{W}{2} = z + \frac{\omega}{2} (e^2
+ p^2) + 2 \lambda e x$ (only at constant $\lambda$).

As the classical system has two degrees of freedom and is not
integrable, its phase space is generally not regular. Some issues
of chaotic dynamics in Dicke model with counterrotating terms were
discussed, e.g., in \cite{KPrants}. At small values of coupling
$\lambda$, and small detunings (i.e., $\omega \approx 1$) the
classical phase space is mostly regular, with a small chaotic
layer around the separatrix of the unperturbed ( $\lambda=0$)
problem. As the value of the coupling is increased, the chaotic
region of the phase space also grows, and near the critical
coupling "global" chaos sets in.

It is interesting to note that in our case, for a passage through
the critical point, chaos is not relevant. Dynamics in a vicinity
of the stable fixed point is regular. In the critical region,
classical trajectories do diverge from each other, however this
dynamics is captured by the Painleve equation. After passing the
critical region, dynamics is again confined in the regular regions
of the phase space (provided the sweep was not too fast). A
passage through the critical region can be analyzed in a similar
way to the models of Section II and III.

Introducing pairs of canonically conjugated variables
$(z,\theta)$, $(I,\phi)$ as

\bea x &=& \sqrt{1-z^2} \cos \theta \nonumber\\
     y &=& \sqrt{1-z^2} \sin \theta \\
     p &=& \sqrt{2I} \cos \phi \nonumber\\
     e &=& \sqrt{2I} \sin \phi, \nonumber
\eea we get a Hamiltonian with two degrees of freedom

\be H = \frac{W}{2}= z + \omega I + 2 \lambda \sqrt{2 I( 1-z^2) }
\cos \theta \sin \phi. \ee

At $\lambda=0$, dynamics is (obviously) integrable, and the system
possesses two classical actions: $I_1=I, \quad I_2 = 1+z$. In a
fully classical approach, the initial state at $\lambda=0$ would
be the equilibrium $z=-1$, $I=0$,  i.e. $I_{1,2}=0$. Sweeping the
value of $\lambda$, the system would stuck in this equilibrium
even when it becomes unstable at $\lambda$ larger than  critical.
In the TWA semiclassical approach, an ensemble of phase points
with initial actions $I_{1,2} \sim \frac{1}{N}$ should be
considered. As a result of sweeping $\lambda$ through the phase
transition, both actions undergo changes which depend on the
sweeping rate $\eps$. Provided the parameter $\eps/ \ln N$ is not
too large (i.e., the sweep is not to fast, {\em or} the number $N$
is large enough), the final dynamics will take place in the {\em
regular} region of the phase space, as shown below. In this
region, it is possible to introduce two classical actions which
depend on initial ones and the sweeping rate. Thus, short-time
dynamics after the passage through the quantum phase transition
can be studied using classical adiabatic theory.

To study the dynamics of the phase transition, we need to obtain
an effective Hamiltonian in the vicinity of the critical region.
We do it through the set of canonical transformations described in
the Appendix C. They bring us to the Hamiltonian \be H = 32(P_1^2
+Q_1^2) + \frac{P^2}{2} + \frac{Q^4}{2} - \eps t \frac{Q^2}{2}.
\ee In the lowest order approximation, the adiabatic invariant
associated with the first pair of variables $P_1,Q_1$ is
conserved, while that associated with another pair of variables
$P,Q$ experiences a jump according to PII.

After calculating this jump (which is equivalent to that found in
previous Section), we consider "final" dynamics at couplings
stronger than the critical. There exists several possible regimes
of motion. Figures 15-17 demonstrate, in particular, three
distinct regimes of the dynamics after the sweep. If the final
coupling is far away from the critical (Fig.15), effective
Hamiltonian is that of two weakly coupled harmonic oscillators.
One can see that the photon number undergoes harmonic oscillations
with almost constant amplitude, which can be taken as a
quantitative measure of the nonadiabaticity. On the other hand, if
the final coupling is close to the critical one, while the
sweeping rate is considerably high, one can observe rapid
deviation between TWA and quantum dynamics due to interference
terms not accounted for by the TWA method in the lowest order
approximation. Trajectories entering chaotic regions of the phase
space also make semiclassical dynamics complicated. However, an
interesting and important regime of motion happens in case the
final coupling is not far away from the critical one (such that
the expansions mentioned above remain valid), while the sweeping
rate is sufficiently slow to leave most of the classical
wavepacket in the regular region of the phase space (Fig.17).
Then, a phase point acquires a classical actions $I_2$ by the time
the pulse $\lambda(t)$ is accomplished and keeps the classical
action $I_1$ to be almost zero ($I_{1,2}$ are now related,
correspondingly, to local variables $(q_1,p_1)$ and $(q_2,p2)$
introduced in the Appendix C rather than to $z$ and $I$ as above).
Considerable magnitude of $I_2$ leads to effective dephasing
between different trajectories, as seen in Fig.17 (effective final
Hamiltonian is not linear). As a result the photon number
undergoes {\em damped} oscillations around certain mean value. We
consider this regime in detail. As a signature of nonadiabaticity,
we take the mean (time-averaged) value of amplitude of
oscillations of number of photons. Faster sweeps lead to a shift
of this value downwards (see Fig.17). As shown in Appendix C, in
close analogy with Section III, after averaging over the initial
TWA distribution one has

\be  \bar{ \langle n \rangle} \approx  4 \delta \lambda_f -
\frac{\langle I \rangle}{4 \sqrt{\delta \lambda_f}}, \label{nbar2}
\ee where the final coupling is $\lambda_f = \frac{1}{2}+ \delta
\lambda_f$  and the mean action of trajectories of the classical
wavepacket is \be \langle I \rangle = \frac{3}{4N} +
\frac{8\sqrt{2} \eps}{\pi} \left( C_{\gamma} + \ln \left( \frac{2N
\eps 8 \sqrt{2}}{\pi} \right) \right). \label{I2final2}
 \ee

This prediction is compared with TWA numerics for $N=500$ and
$\delta \lambda_f =0.03$ in Fig.17. Remarkable coincidence can be
seen at relatively slow sweeps.

\section{Conclusions}

We provided a method to treat the dynamics of quantum phase
transitions in systems whose classical counterparts have a few
degrees of freedom. The initial state of a quantum system (which
is chosen to be close to its ground state) is represented by an
ensemble of classical phase points initially concentrated in a
$\frac{1}{N}-$vicinity of a stable fixed point. By sweeping a
parameter, the quantum system is transferred through a quantum
phase transition, which in the underlying classical system
corresponds to a bifurcation. Second-order QPT usually corresponds
 to a pitchfork bifurcation, while first-order QPT to saddle-centre
bifurcation. Magnitudes of quantum observables are obtained by
averaging over the classical distribution. From the properties of
the Wigner function it follows that the initial angle of the phase
points in the ensemble (i.e., the phase canonically conjugated to
an action variable) is uniformly distributed on ($0,2\pi$).  As a
result of the passage through a bifurcation, the canonical action
undergoes a phase-dependent jump.  As a function of the sweeping
rate and initial action, this jump for a pitchfork bifurcation has
very peculiar properties.

A central idea of the method is as follows:  far from the
bifurcation, classical actions are approximately conserved (due to
adiabatic invariance of action). In the vicinity of a bifurcation,
the system can be brought to a "normal" form: the first Painleve
equation for the saddle-center bifurcation, and the second
Painleve equation for the pitchfork bifurcation. Change in the
action is determined by these effective equations. The method
provides an asymptotically exact theory, i.e. coefficients of the
power-laws derived, etc.,  are asymptotically {\em exact}. To be
specific, we considered here several particular models: Dicke
models with and without counterrotating terms, and the
Lipkin-Meshkov-Glick model.


The results are as follows.

(i) The most important results for the Dicke model without
counterrotating terms: (i) the linear-logarithmic  power-law for
the forward sweep, and the linear power-law with the coefficient
$\frac{\ln 2}{\pi}$ for the inverse sweep. Properties of the
initial Wigner distribution are such that additional trajectories
not described by the PII equation appear, which modify the
coefficients of the power-law in the forward sweep.

(ii) The most important results for the Dicke model with
counterrotating terms are as follows. Even though it is usually
assumed that superradiant phase corresponds to chaotic classical
dynamics, i.e. the bifurcation at critical coupling leads to
chaotic phase space, we found that close to the new equilibria
(emerging in the superradiant phase) phase space is regular; it is
possible to introduce two classical actions there. Now, in the
limit of large $N$ initial classical wavepacket is highly
localized near the equilibrium; therefore at not very fast sweeps
it will remain in the regular area of the phase space after the
sweep. One may ask what changes in the classical actions will be.
We found the dynamics of passage through bifurcation happens in
the way where, although there are two pairs of canonical variables
those frequencies are of the same order far from bifurcation, in
the vicinity of the bifurcation a separation of time-scales
happens and by a suitable transformation it is possible to define
a pair of (slow) variables those dynamics is described by PII and
another pair of (fast) variables which conserves its classical
action. Briefly, passage through the bifurcation in the
multidimensional system happens effectively in a 'one-dimensional'
way.

(iii) The Dicke model with large discrepancy of frequencies and
large dissipation, relevant to the \cite{Esslinger} experiment,
can be approximated by the LMG model \cite{Keeling2}. The passage
through a bifurcation there is also described by PII. What is
exciting in this model is that all trajectories from the initial
distribution are described by PII.

An important question is how to detect the final distributions
experimentally. While in the model of Section II one can directly
measure the final distributions of number of molecules,  in the
case of LMG model (Section III) and sweep to $\gamma=0$ the
situation is more complicated. A possible scenario would involve
measuring the time-averaged dispersion of $J_z$ after the sweep,
which will depend on the distribution of the final action due to
the nonlinearity of the effective Morse oscillator. In the case of
Model of Section IV  one can monitor (for instance) a shift in the
time-averaged value of the photon number as a function of the
sweeping rate.

\section*{ Acknowledgements} This work was supported by the Academy
of Finland (Projects No. 213362, No. 217043, and No. 210953) and
EuroQUAM/FerMix, and conducted as a part of a EURYI scheme grant
(see www.esf.org/euryi). A.P.I. was partly supported by RFBR
09-01-00333.  We are grateful to A.I. Neishtadt, A.Polkovnikov,
V.Gurarie, A.B.Klimov, L.Plimak for illuminating discussions.

\section*{Appendix A: Wigner function distributions}

Very useful approximate expressions for SU(2) Wigner functions of
$J_z$ operator eigenstates were obtained in
\cite{ASchleich,Klimov}. In the limit of large dimension of
representation $J \gg 1$,  result of \cite{Klimov} for $J_z$
eigenstate $|k,J \rangle$ is

\be W_k(\theta, \phi) \approx (-1)^{J+k} d^J_{kk}(2 \theta) [1+
\frac{k}{J} \cos{\theta}] \frac{N+1}{2} , \ee where $d^J_{kk}$ is
the (small) Wigner d-function, which is related to Jacobi
polynomials.

\begin{figure}
\includegraphics[width=80mm]{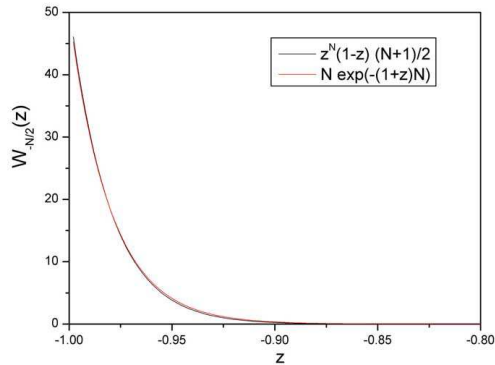}
\caption{The Wigner function of the state
$|\frac{N}{2},-\frac{N}{2}\rangle$.  Here we compare two functions
approximating the exact SU(2) Wigner function,
$z^N(1-z)\frac{(N+1)}{2}$, and $N \exp[-(1+z)N]$, for $N=50$. Note
that the two curves are almost indistinguishable on that scale.}
\label{FA10}
\end{figure}

For the ground state $k=-J=-\frac{N}{2}$  one gets \bea W_{-J}
(\theta, \phi) &=&  d^J_{-J,-J} (2 \theta) [1-\cos
\theta]\frac{N+1}{2} \\ &=& (\cos \theta)^{2J} [1-\cos
\theta]\frac{N+1}{2} = z^N (1-z)\frac{N+1}{2}. \nonumber\eea This
function behaves as an exponential function near $\theta=\pi
(z=-1) $
 (see Fig.\ref{FA10}): \be
 W_{-J} (z, \phi)|_{z \approx -1} \approx  N\exp[-N(1+z)].
\ee

\section*{Appendix B: Poincare surfaces of section for the  Dicke model with
counterrotating terms}

Below we present Poincare surfaces of section of the Dicke model
with counterrotating terms demonstrating: (i) chaotization of
large part of the phase space  (ii) remaining regular parts near
stable equilibria.
\begin{figure*}
\includegraphics[width=140mm]{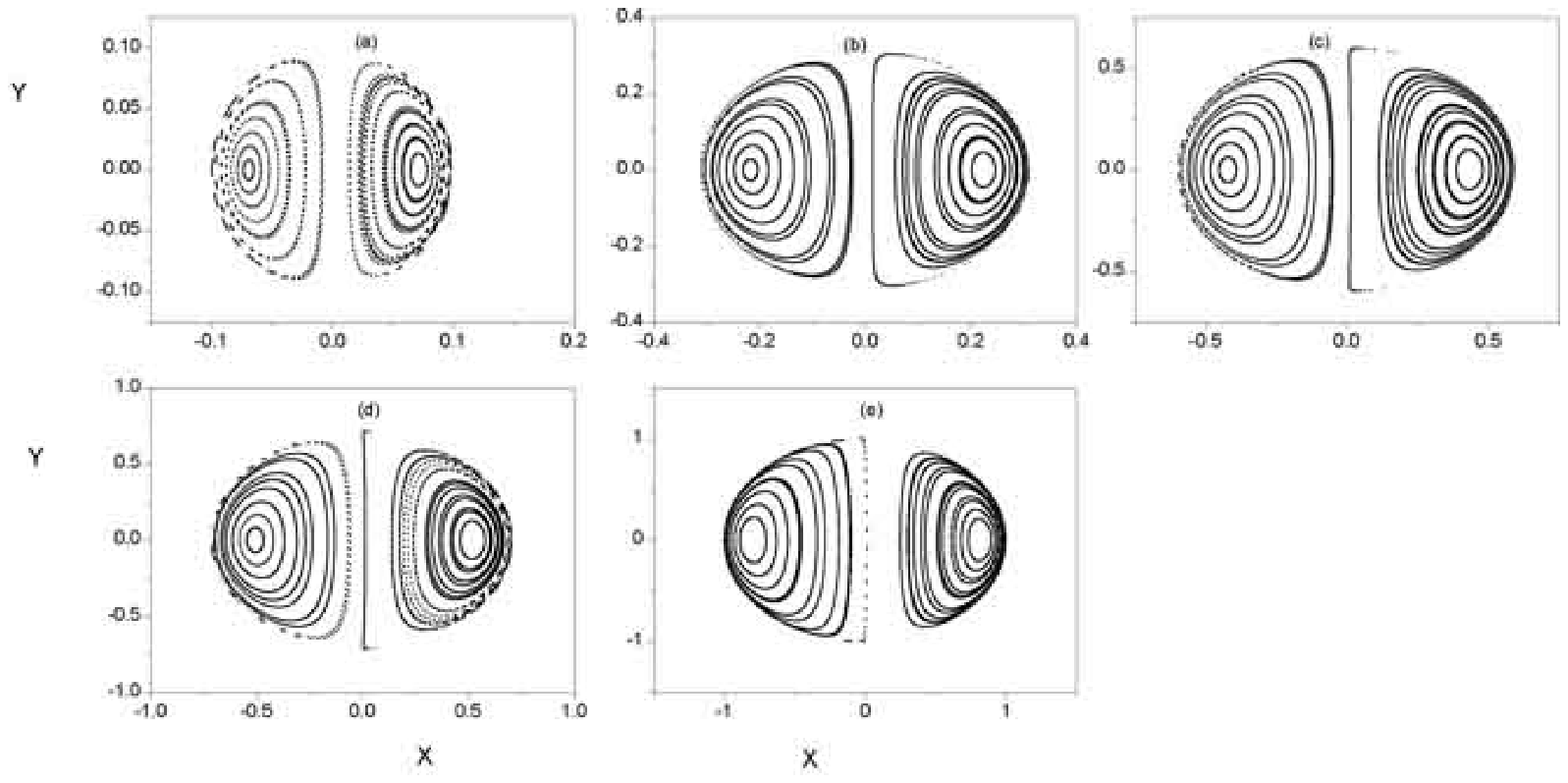}
\caption{Poincare surfaces of section, weak coupling
$\lambda=0.02$. Different energy surfaces are shown:
h=-0.995,-0.95,-0.8,-0.7,0.0 for (a-e), correspondingly. }
\includegraphics[width=140mm]{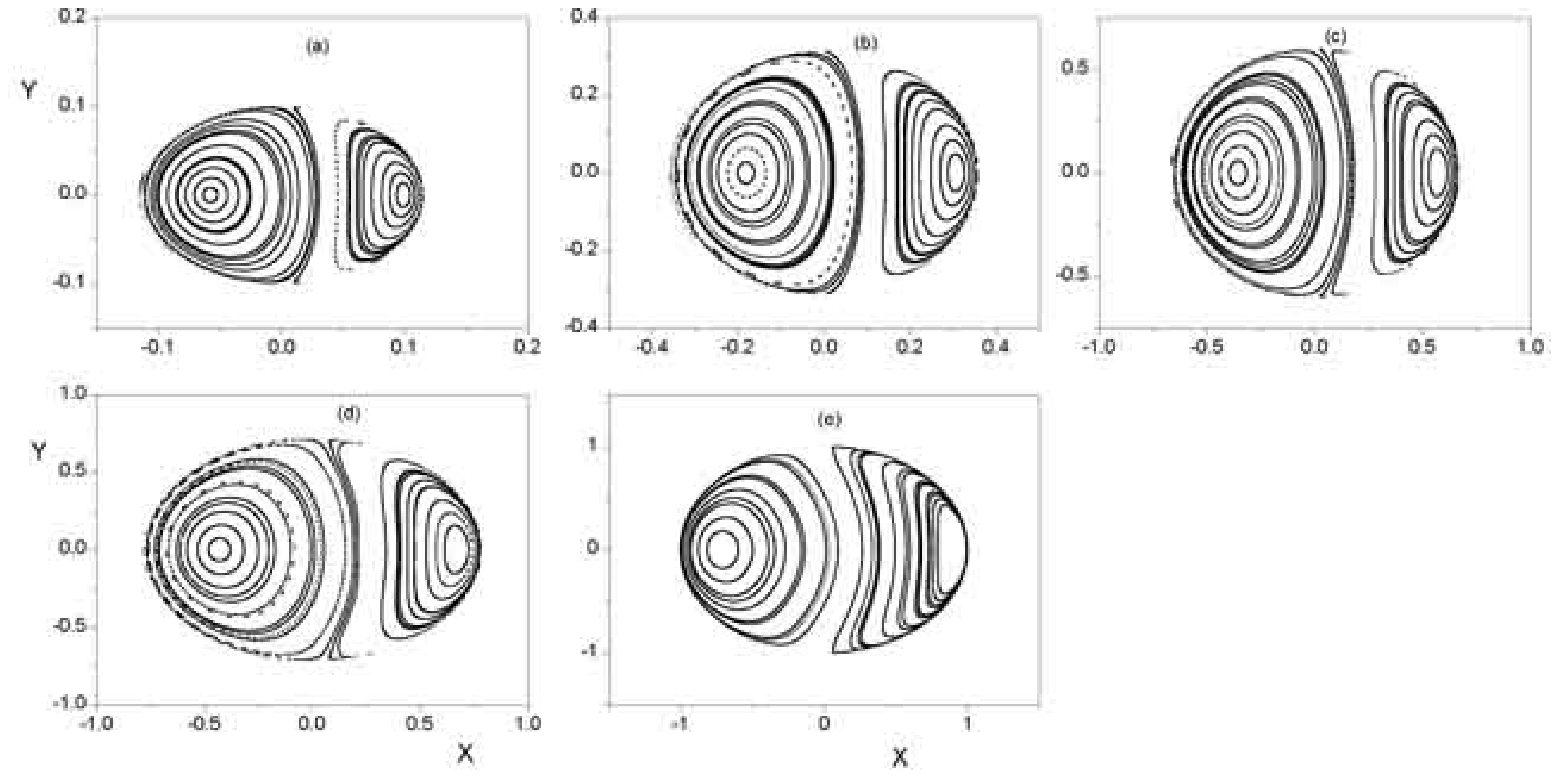}
\caption{Poincare surfaces of section, $\lambda=0.25$.}
\includegraphics[width=130mm]{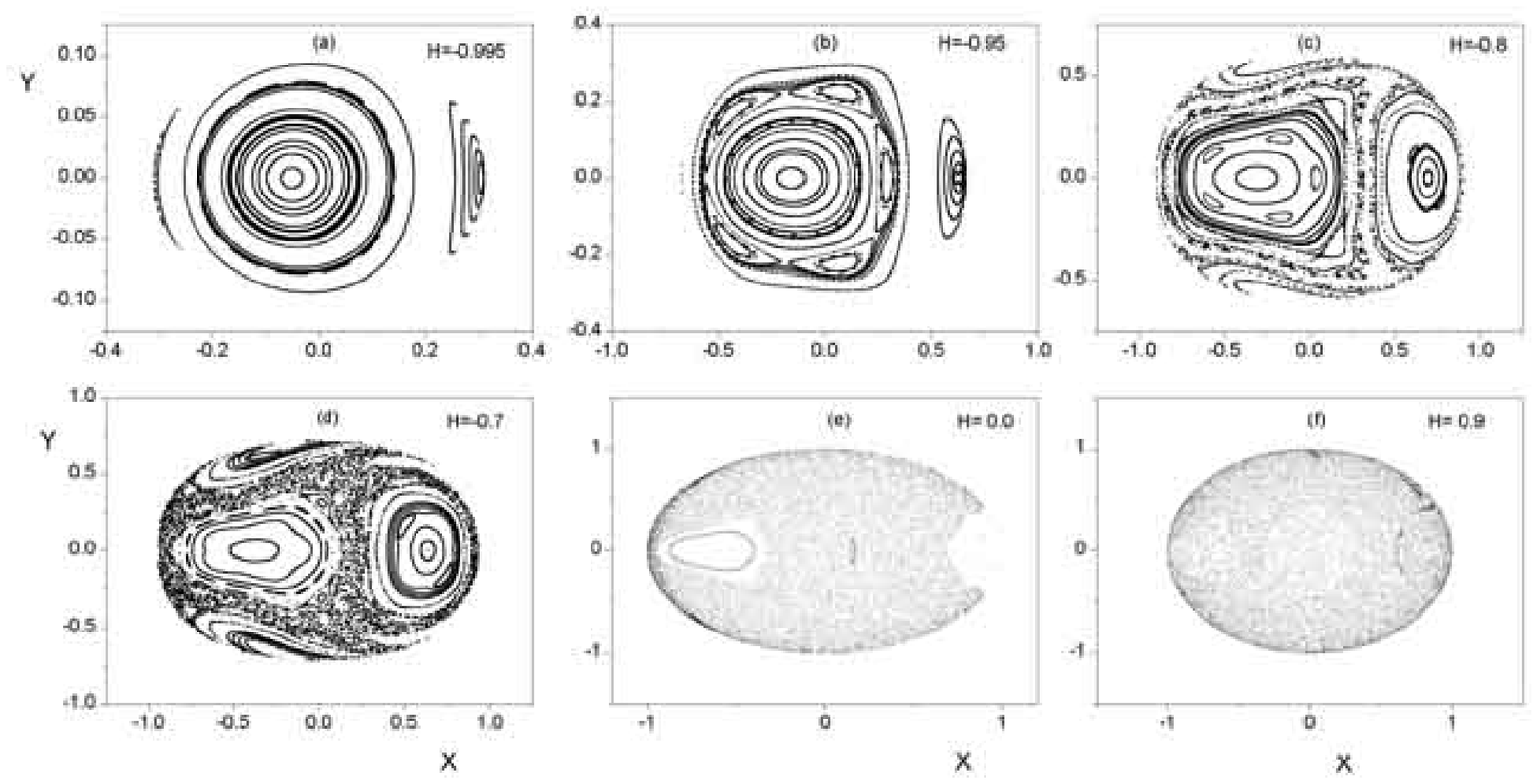}
\caption{Poincare surfaces of section, $\lambda=0.48$ (close to
critical).}
\end{figure*}
\begin{figure*}
\includegraphics[width=130mm]{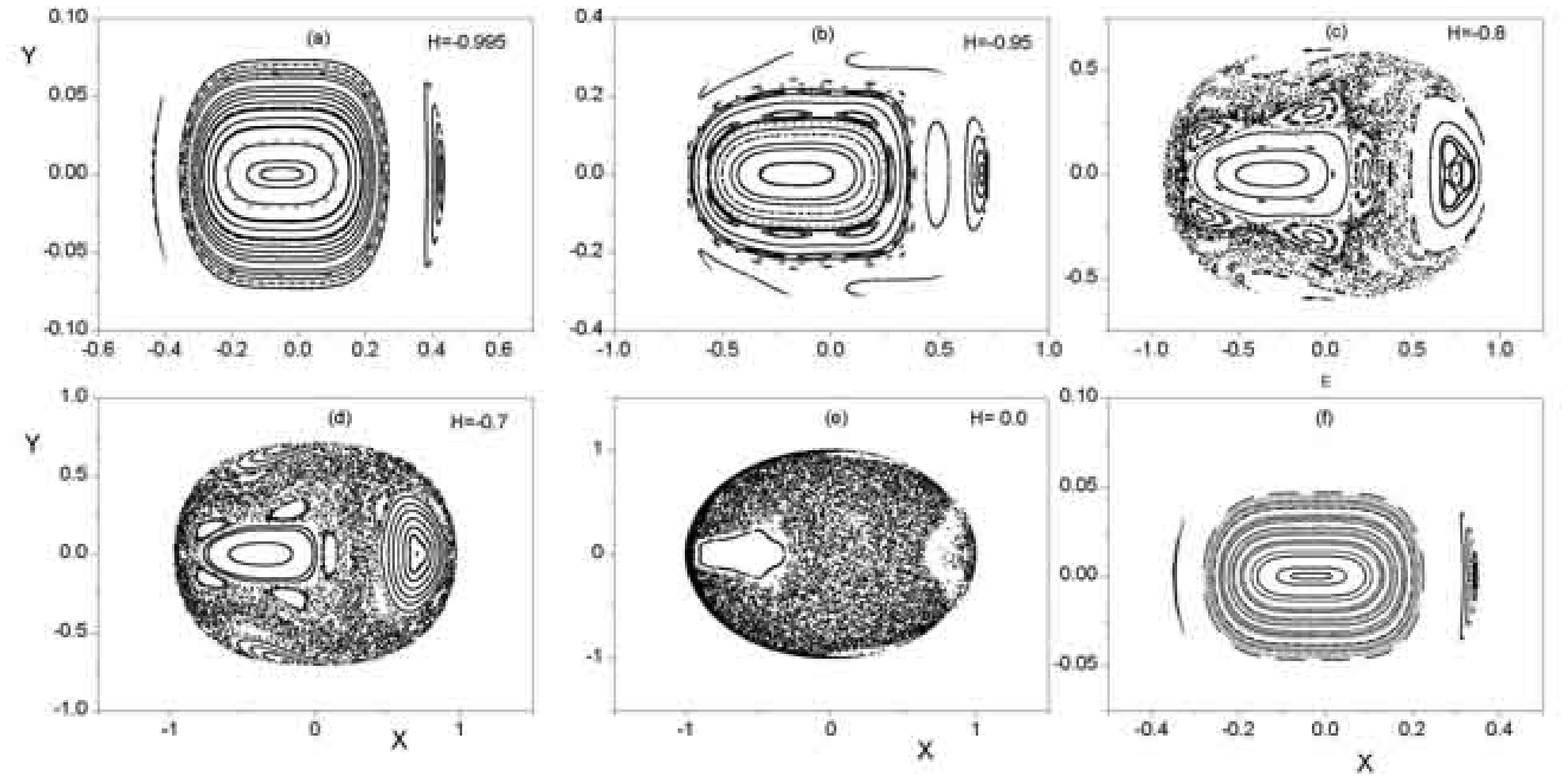}
\caption{Poincare surfaces of section, $\lambda=0.5$.}
\includegraphics[width=130mm]{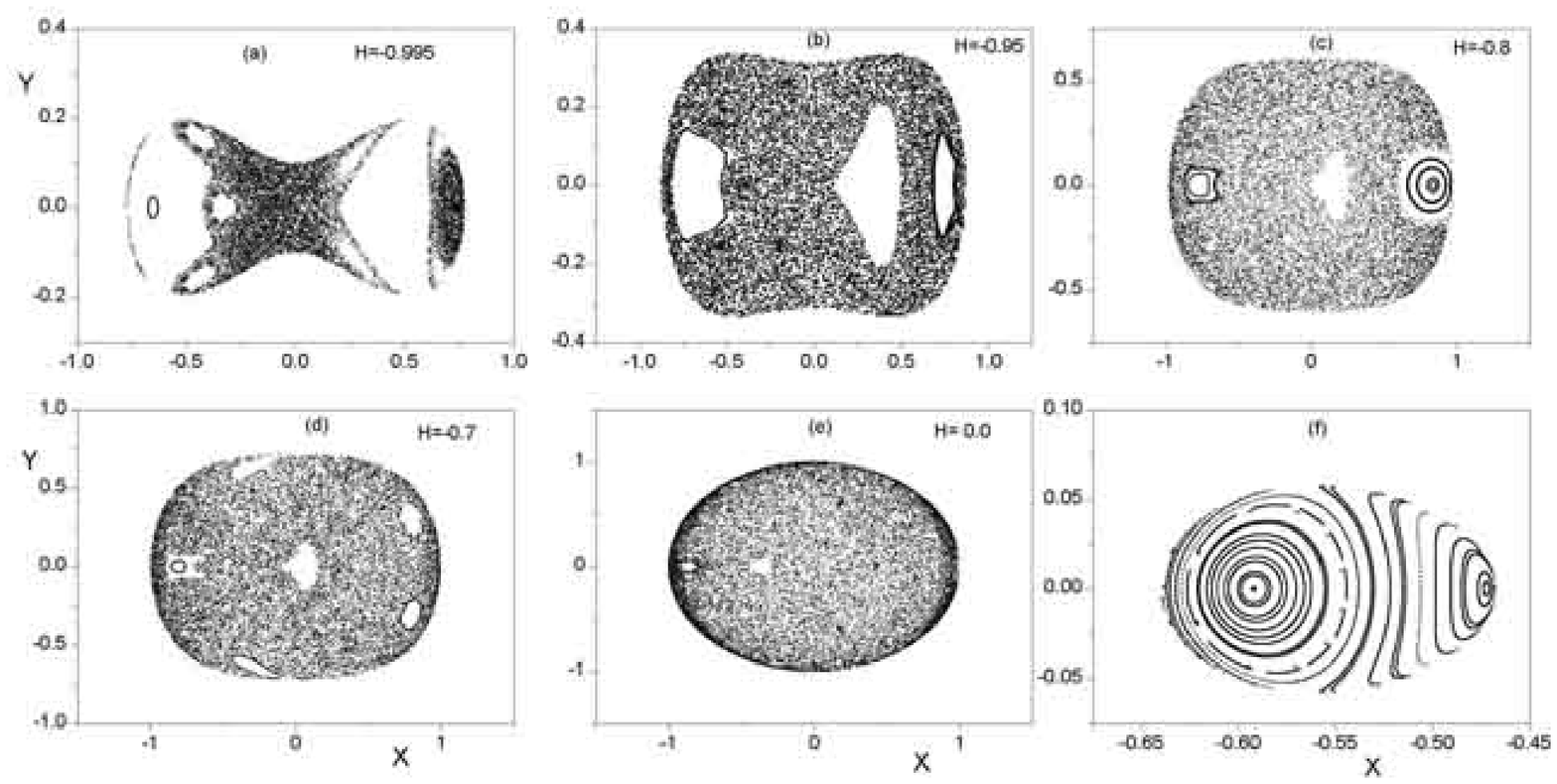}
\caption{Poincare surfaces of section, $\lambda=0.55$. Note that
Fig. (f) corresponds to an energy level close to that of  new
equilibria. Poincare surface of section is regular there. }
\includegraphics[width=130mm]{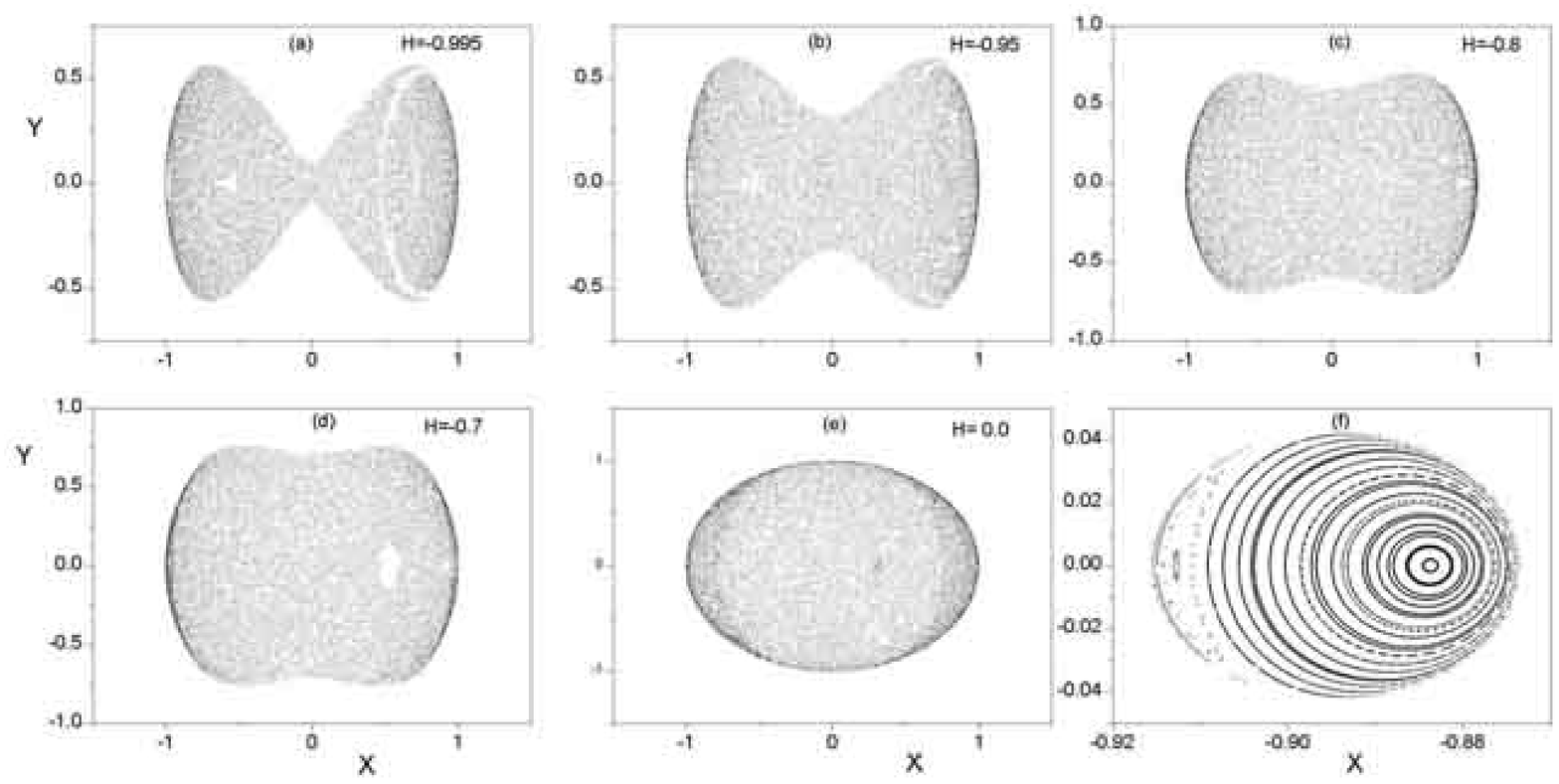}
\caption{Poincare surfaces of section, $\lambda=0.75$. Note that
Fig. (f) corresponds to an energy level close to that of  new
equilibria. Poincare surface of section is regular there.}
\end{figure*}

\newpage
\section*{Appendix C: mapping to Painleve equation in the Dicke model with counterrotating terms}

Let us firstly do a canonical transformation to variables
$(P_Y,Y), (P_X,X)$: $P_Y = \sqrt{2(1+z)} \sin \theta, \quad Y = -
\sqrt{2 (1+z)} \cos \theta$, $ P_X=p, \quad X=e $. In the new
variables $$ H = \frac{\omega}{2}(P_X^2 + X^2) + \frac{1}{2}(P_Y^2
+ Y^2) - 2 \lambda XY \sqrt{1- \left(\frac{P_Y^2+Y^2}{4}\right)}
$$ Before the bifurcation, the equilibrium is in the origin. Near
the origin, expanding the term with a square root in $H$ in
Taylor' series, we obtain approximately the equations of motion
\bea \dot{P}_X &=& - \omega X + 2 \lambda Y (1-
\frac{P_Y^2+Y^2}{8})
\nonumber\\
\dot{X} &=& \omega P_X  \\
\dot{P}_Y &=&  -Y + 2 \lambda X (1- \frac{P_Y^2}{8} -\frac{3
Y^2}{8})  \nonumber\\
\dot{Y} &=& P_Y + \lambda XY \frac{P_Y}{2}. \nonumber \eea
Analyzing fixed points of the Hamiltonian, it is not difficult to
 see that at $\lambda=\lambda_{crit} = \frac{1}{2}$ a bifurcation
happens and new equilibria at $Y= \sqrt{2 \left[1-
\frac{\omega}{(2 \lambda)^2} \right]}, \quad P_Y=0, \quad P_X=0, X
= \pm \frac{2 \lambda}{\omega}
\sqrt{1-\frac{\omega^2}{2(\lambda)^4}}$ emerge.

From now on, consider the case $\omega=1$ for simplicity.
Introducing the new variables $\tilde{P}_X = \sqrt{\omega}P_X,
\quad \tilde{X}=X/\sqrt{\omega}$, and making a rotation

\be \tilde{X}=(q_1+q_2)/\sqrt{2}, \quad Y = (q_2-q_1)/\sqrt{2},
\ee we obtain \bea H &=& \nonumber\\ \frac{1}{2}(p_1^2 &+& p_2^2
+q_1^2 +q_2^2) + \lambda
(q_1^2 - q_2^2 ) +  \frac{\lambda}{16}(q_2^4 - q_1^4) + \nonumber\\
\frac{\lambda}{16}(q_2^2 &-& q_1^2)(p_1^2 + p_2^2 - 2 (p_1 p_2
+q_1 q_2 )).  \eea In a vicinity of the critical point
$\lambda=\frac{1}{2}$, the effective Hamiltonian is

\be H= \frac{p_1^2}{2} + \frac{q_1^2}{2}(1 + 2 \lambda) +
\frac{p_2^2}{2} + \frac{q_2^2}{2}(1-2\lambda) +
\frac{\lambda}{16}q_2^4. \label{Hbare} \ee Shifting the origin of
time, approximating the change in $\lambda$ by a linear law in the
vicinity of the critical value $\lambda = \frac{1}{2}+ \eps t$,
and neglecting time dependence in the coefficient $1+2 \lambda $,
we obtain: \be H= \frac{p_1^2}{2} + q_1^2 + \frac{p_2^2}{2} -
\frac{q_2^2}{2}(2\eps t) + \frac{1}{32}q_2^4. \ee Neglecting the
time dependence in the coefficient $1+2 \lambda $ does not prevent
one from obtaining correct lowest-order approximation of
destruction of adiabaticity, since the change in the adiabatic
invariants mostly happens around the time of bifurcation.
Introducing new variables and rescaling the Hamiltonian as

$ p_2 = 8 P, \quad p_1 = 8 P_1, \quad q_2 = 4 \sqrt{2} Q, \quad H
= 64H', $ we obtain the Hamiltonian

\be H' = 32(P_1^2 +Q_1^2) + \frac{P^2}{2} + \frac{Q^4}{2} - \eps t
\frac{Q^2}{2}. \label{Hprime} \ee

Sweep through the bifurcation results in change of the adiabatic
invariant analogous to that considered in Sections II-III.

Let us calculate quantitatively  effects of non-adiabaticity
averaged over the classical ensemble. We see in Fig.17 that after
the sweep the photon number oscillates around a (time-averaged)
mean value $n_0(\eps)=\bar{\langle n \rangle}$.  Increasing the
sweeping rate $\eps$ leads to shift of the mean value $n_0$. The
magnitude of this shift can be used to quantify the deviation from
adiabaticity in the quasiclassical ensemble. Indeed, consider a
single trajectory of the Hamiltonian  $h=\frac{p^2}{2} -A
\frac{q^2}{2} + B \frac{q^4}{4}$ (the effective final Hamiltonian
after the sweep). It is easy to see that the magnitude of $\bar n
= \frac{1}{4}(\overline{ p^2+q^2}) $ in the first approximation
linearly depends on the action of the trajectory $I$:  $ \bar n =
\alpha - \beta I $.  Explicit values of the coefficients
$\alpha,\beta$ can be obtained from the following estimates:

(i) the relation between the action $I$ and the energy $h$ near a
stable fixed point:

\bea I &=& \frac{1}{\pi} \int_{q_1}^{q_2} p dq =
 \frac{1}{\pi} \int_{x_1}^{x_2} \sqrt{2h
+Ax -\frac{B x^2}{2}}\frac{dx}{2\sqrt{x}} \\
&=&\frac{1}{3\pi}\sqrt{\frac{B x_2}{2}} \left[(x_2 +
x_1)\mbox{E}(m) - 2x_1 \mbox{K}(m) \right] \approx
\frac{x^2}{4\sqrt{2AB^2}}, \nonumber\eea

where \be  x_{1,2}=q_{1,2}^2 = \frac{A \mp x}{B}, \quad
x=\sqrt{A^2 + 4B h}, \ee and $m = \frac{x_2-x_1}{x_2}$.

(ii) time-averaged magnitudes of $p^2,q^2$:

\bea \overline{q^2} &=& \frac{1}{T} \int q \frac{dq}{\dot{q}} =
\frac{1}{T} \int_{x_1}^{x_2} \frac{dx \sqrt{x}}{\sqrt{2h + Ax -
\frac{B}{2}x^2}} \\&=& x_2 \frac{\mbox{E}(m)}{\mbox{K}(m)} \approx
\frac{A}{B} - \frac{x^2}{4AB}, \nonumber\\
 \overline{p^2} &=& \frac{1}{T} \int_{x_1}^{x_2}dx
\frac{\sqrt{2h + Ax - \frac{B}{2}x^2}}{\sqrt{x}}   \approx
 \frac{x^2}{4B}, \eea

where the period $T = \int_{x_1}^{x_2} \frac{dx}{\sqrt{x}\sqrt{2h
+ Ax - \frac{B}{2}x^2}} = \frac{2}{\sqrt{x_2}}\sqrt{\frac{2}{B}
\mbox{K}(m)}$.

We have therefore

\bea \bar n &=& \frac{1}{4}(\overline{p^2+q^2}) \approx
\frac{1}{4B}\left(A + x^2(\frac{1}{4}- \frac{1}{4A} )\right)
\nonumber\\ &=& \frac{1}{4B}\left(A + I B \sqrt{2A} (1-
\frac{1}{A} )\right). \eea One can note that  $\overline n$ is
linearly proportional to the excess energy of a trajectory (i.e.
the difference between the energy of the final solution and the
adiabatic solution), which is often used for quantifying
non-adiabaticity.

After averaging over the classical TWA ensemble, \be  \bar{
\langle n \rangle} = \frac{A}{4B} + \langle I \rangle
\sqrt{\frac{A}{8}}\left(1-\frac{1}{A} \right), \ee

where $\langle I \rangle$ is the mean value of the adiabatic
invariant associated with the variables $p_2,q_2$. We see that the
($\eps-$dependent) shift of $\overline n$ gives a convenient
quantitative measure of the nonadiabaticity.  If the sweep of
$\lambda$ ends at some magnitude $\lambda_f$ being close to the
critical one $\lambda_f = \frac{1}{2} + \delta \lambda_f$, $\delta
\lambda_f \ll 1$, we have

\be  \bar{ \langle n \rangle} \approx \frac{A}{4B} - \langle I
\rangle \frac{1}{\sqrt{8A}} = 4 \delta \lambda_f - \frac{\langle I
\rangle}{4 \sqrt{\delta \lambda_f}}. \label{nbar} \ee

Let us find now the $\eps-$dependence of $\langle I \rangle$.

The action $I_2$ of the Hamiltonian (\ref{Hbare}) is related to
the action $I_{PQ}$ of the Hamiltonian (\ref{Hprime}) as
$I_2=32\sqrt{2} I_{PQ}.$

The final action $I_2$ of a trajectory from the classical TWA
ensemble is related to its initial action as

\be I_2^+  \approx \frac{3}{4} I_2^- - \frac{8\sqrt{2} \eps}{\pi}
\ln \left( \frac{2 \pi I_2^-}{32 \sqrt{2} \eps} \right). \ee

Averaging over the TWA ensemble leads to

\be \langle I_2 \rangle = \frac{3}{4N} + \frac{8\sqrt{2}
\eps}{\pi} \left( C_{\gamma} + \ln \left( \frac{2N \eps 8
\sqrt{2}}{\pi} \right) \right). \label{I2final}
 \ee

Substituting this expression into Eq.\ref{nbar} in place of
$\langle I \rangle $, we obtain the theoretical prediction for the
shift of the mean photon number $n$. Note that we do not consider
$1/N$ differences between quantum observables and semiclassical
TWA ones, but only present the TWA-Painleve prediction for the
semiclassical ensemble. This prediction is compared with TWA
numerics for $N=500$ and $\delta \lambda_f =0.03$ in Fig.17.
Remarkable coincidence can be seen at slow sweeps.




\begin{thebibliography}{10}
\bibitem{explanation} Formally, classical system is deterministic: initial conditions and
equations of motion are fixed. Nevertheless, fate of an individual
classical trajectory is highly dependent on the initial phase
(being uniformly distributed), and therefore is "quasi" random.
Probabilistic phenomena in dynamical systems were discussed, e.g.,
in \cite{probabilistic,AKN}.
\end{thebibliography}
\end{document}